\documentclass[11pt,draftcls,onecolumn,journal]{IEEEtran}
\usepackage{bbm}
\usepackage{amsmath,amsfonts}
\usepackage{algorithmic}
\usepackage{algorithm}
\usepackage{array}
\usepackage{textcomp}
\usepackage{stfloats}
\usepackage{url}
\usepackage{verbatim}
\usepackage{graphicx}
\usepackage{cite}
\usepackage{hyperref}
\usepackage{booktabs}
\usepackage{bm}
\usepackage[most]{tcolorbox}
\usepackage{capt-of}
\usepackage{makecell}
\usepackage{color, colortbl}
\usepackage{pdflscape}

\usepackage{multirow}
\usepackage{arydshln}
\usepackage{dsfont}
\usepackage{amssymb}
\usepackage{enumitem} 
\usepackage{tikz}
\usetikzlibrary{tikzmark}
\usetikzlibrary{positioning, calc}
\usetikzlibrary{matrix}
\usepackage{array}
\usetikzlibrary{shadows}
\usetikzlibrary{arrows, fit, positioning, shapes,backgrounds}
\pgfdeclarelayer{background}
\pgfdeclarelayer{foreground}
\pgfsetlayers{background,main,foreground}
\tikzstyle{materia}=[draw, fill=uzh@lightblue, text width=6.0em, text centered,
  minimum height=1.5em,drop shadow]
\tikzstyle{practica} = [materia, text width=20em, minimum width=20em,
  minimum height=3em, rounded corners, drop shadow]
\tikzstyle{materia2}=[draw, fill=red, text width=6.0em, text centered,
  minimum height=1.5em,drop shadow]
\tikzstyle{practica2} = [materia2, text width=8em, minimum width=10em,
  minimum height=3em, rounded corners, drop shadow]
  
\tikzstyle{texto} = [above, text width=6em, text centered]
\tikzstyle{linepart} = [draw, thick, color=black!50, -latex, dashed]
\tikzstyle{line} = [draw, thick, color=black!50, -latex]
\tikzstyle{ur}=[draw, text centered, minimum height=0.01em]

\usetikzlibrary{bayesnet}

\usetikzlibrary{shapes}
\usepackage{ulem}

\tikzset{
    targetbox/.style={
        draw,
        dashed,
        rounded corners,
        fill=yellow!20,
        minimum width=3cm,
        minimum height=0.9cm,
        align=center
    }
}

\usepackage{multirow}
\usepackage{subcaption}

\newtheorem{lemma}{Lemma}
\newtheorem{definition}{Definition}

\definecolor{Gray}{gray}{0.9}
\newcolumntype{H}{>{\setbox0=\hbox\bgroup}c<{\egroup}@{}}
\newcolumntype{C}{>{\centering\let\newline\\\arraybackslash\hspace{0pt}}m{4cm}}

\newcommand{\mv}{\bm{m}}
\newcommand{\nv}{\bm{n}}

\newcommand{\uv}{\bm{u}}

\newcommand{\wv}{\bm{w}}
\newcommand{\xv}{\bm{x}}
\newcommand{\yv}{\bm{y}}
\newcommand{\zv}{\bm{z}}
\newcommand{\zerov}{\bm{0}}
\newcommand{\onev}{\bm{1}}

\newcommand{\Ab}{\bm{A}}

\newcommand{\Ib}{\bm{I}}

\newcommand{\Lb}{\bm{L}}
\newcommand{\Mb}{\bm{M}}
\newcommand{\Nb}{\bm{N}}

\newcommand{\Rb}{\bm{R}}
\newcommand{\Sb}{\bm{S}}

\newcommand{\Ub}{\bm{U}}
\newcommand{\Vb}{\bm{V}}
\newcommand{\Wb}{\bm{W}}
\newcommand{\Xb}{\bm{X}}
\newcommand{\Yb}{\bm{Y}}
\newcommand{\Zb}{\bm{Z}}

\newcommand{\thev}{\bm{\theta}}
\newcommand{\muv}{\bm{\mu}}

\newcommand{\epv}{\bm{\epsilon}}

\newcommand{\psiv}{\bm{\psi}}

\newcommand{\gamv}{\bm{\gamma}}

\newcommand{\phiv}{\bm{\phi}}

\newcommand{\Lamb}{\bm{\Lambda}}

\newcommand{\Sigmab}{\bm{\Sigma}}

\newcommand{\Real}{\mathbb{R}}

\newcommand{\Nc}{{\cal N}}

\newcommand{\Sc}{{\cal S}}

\newcommand{\norm}{\Vert}
\newcommand\normop[1]{\left\lVert#1\right\rVert}

\newcommand{\lone}{$L_{1}$}
\newcommand{\ltwo}{$L_{2}$}
\newcommand{\lp}{$L_{p}$}

\newcommand{\minn}{\mathop{ \mathsf{min}}}
\newcommand{\maxx}{\mathop{ \mathsf{max}}}
\newcommand{\argminn}{\mathop{ \mathsf{argmin}}}
\newcommand{\argmaxx}{\mathop{ \mathsf{argmax}}}
\newcommand{\tr}{\mathop{ \mathrm{tr}}}

\newcommand{\diag}{\mathrm{diag}}

\newcommand{\rank}{\mathrm{rank}}
\newcommand{\isdef}{\mathrel{\overset{\makebox[0pt]{\mbox{\normalfont\tiny\sffamily d}}}{=}}}

\newcommand{\PSD}{\Sc_p^{+}}

\newcommand{\Xtrain}{\bm{X}_{\textrm{train}}}
\newcommand{\Xtraino}{\bm{X}_{\textrm{train},\sf o}}
\newcommand{\xnew}{\bm{x}_{\textrm{new}}}
\newcommand{\xnewo}{\bm{x}_{\textrm{new}, \sf o}}
\newcommand{\ytrain}{\bm{y}_{\textrm{train}}}
\newcommand{\ynew}{{y}_{\textrm{new}}}

\newcommand{\xbo}{\bm{x}_{\sf o}}
\newcommand{\xbm}{\bm{x}_{\sf m}}

\newcommand{\Xbo}{\bm{X}_{\sf o}}
\newcommand{\Xbm}{\bm{X}_{\sf m}}

\newcommand{\subto}{\mathsf{s.\,t.}}

 \def\T{{\mkern-2mu\mathsf{T}}}
\begin{document}

\title{Missing Data in Signal Processing \&  Machine Learning:  Models, Methods \& Modern Approaches}

\author{A. Hippert-Ferrer, A. Sportisse, A. Javaheri,  M. N. El Korso, D. P. Palomar
}

\markboth{IEEE Signal Processing Magazine,~Vol.~XX, No.~XX, July~2024}%
{\MakeLowercase{\textit{et al.}}: Author Guidelines for Special Issue Articles of IEEE SPM}

\maketitle


\color{black}
\section*{Introduction}

Missing data is a pervasive challenge across scientific disciplines, occurring whenever observations are partially unavailable due to sensor failures, measurement limitations, human error, or natural phenomena. In signal processing (SP) and machine learning (ML), incomplete data can severely compromise analysis, distort parameter estimates, and undermine predictive performance. 
From sensor networks and radar systems to medical imaging and financial time series, practitioners across domains must routinely grapple with the question: \textit{How should we handle missing data?} For example:
\begin{itemize}   
    \item In {biomedical signal processing}, equipment malfunctions frequently cause missing recordings. For instance, electroencephalography (EEG) signals often contain faulty measurements due to electrode disconnections or motion artifacts, prompting analysts to discard affected segments and creating gaps in the data.
    
    \item In {image processing}, images must often be restored from partial observations caused by distortions such as scratches, noise, or occlusions. In remote sensing, for example, optical sensors are limited by cloud cover and prone to hardware failures, such as the well-known malfunction of the Landsat ETM+ scan-line corrector which produced images with systematic gaps.
    
    \item In {sensor array processing}, multiple antennas record signals from a small number of sources, and measurements may be lost during transmission or due to sensor failures. The resulting data matrix exhibits a low-rank structure, as the number of underlying sources is typically much smaller than the number of sensors.
    
    \item In {financial data analysis}, traded assets often exhibit missing values due to non-trading periods, lack of liquidity, or data handling errors. Accurate imputation is critical for applications such as portfolio optimization, backtesting, and risk modeling.

    \item In {recommender systems}, the Netflix Prize dataset (released in 2006) comprised ratings from approximately 480K users over 18K movies, yet only about 1\% of entries were observed: the competition challenged participants to predict missing ratings from this extremely sparse matrix.
    
    \item In {graph-structured data}, signals defined on networks, such as temperature measurements in sensor networks, user interactions in social networks, or traffic patterns in transportation systems, frequently contain missing entries due to device malfunctions or privacy constraints.
\end{itemize}

\color{black}

Indeed, the challenge of extracting information from incomplete observations is not new. Early statistical work by Wilks in 1932 addressed parameter estimation for bivariate normal distributions with missing values, motivated by incomplete questionnaires in social sciences and government statistics. Lord extended this framework to multivariate settings in 1955 \cite{lord1955}. However, since the early 1970's, the field truly flourished following Rubin's seminal work, which formalized the theoretical foundations of missing-data mechanisms, i.e., the probabilistic relationship between missingness and data values \cite{rubin1976}. This framework established that different causes of missingness require fundamentally different analytical approaches. Combined with advances in computational capacity, Rubin's theory catalyzed major methodological developments spanning statistical inference, data analysis, and machine learning. In the past three decades, the SP and ML communities have witnessed tremendous progress in adapting classical techniques to handle incomplete signals and in developing entirely new methodologies in core domains, including compressive sensing, parameter estimation, graph signal processing, subspace tracking, filtering, regression, classification, and time series analysis. 

Practitioners typically face three fundamental tasks when confronting missing data: \textit{imputing} missing entries to obtain complete datasets, \textit{estimating} parameters or signals from incomplete observations, and \textit{learning} predictive models despite missing values. Each of these analytical tasks presents distinct challenges and has inspired specialized methodologies; yet, these areas are deeply interconnected.

\textit{Scope and contributions of this tutorial.} 
This article aims to provide practitioners with comprehensive, accessible tools for dealing with missing data in modern SP/ML applications. Rather than claiming exhaustive coverage, we focus on both classical and recent techniques tailored to the needs of the SP/ML community. Our tutorial makes three primary contributions:

First, we provide a clear, unified framework for organizing missing-data strategies around three fundamental tasks: \textit{imputation} (replacing missing entries with predicted values), \textit{estimation} (inferring unknown parameters or signals from incomplete data), and \textit{learning} (predicting target values from incomplete observations or using generative (deep)-learning approaches to impute missing values). While these categories are permeable and often intertwined, this organization clarifies the landscape and helps practitioners identify appropriate methods for their applications.

Second, we highlight major methodological advances through detailed case studies grounded in real-world scenarios, demonstrating how contemporary techniques address imputation, estimation, and learning tasks in challenging settings involving irregular, high-dimensional, structured data, and complex missingness mechanisms.

\begin{table}[t]
\arrayrulecolor{black} 
\hspace*{-20pt}
    \centering
    \begin{tabular}{c|c|c|c}
        Framework & Formulation & Algorithms / Approaches & Reference in the text   \\ \hline
         \multirow{2}{*}{Imputation} & \begin{tabular}[c]{@{}c@{}} Imputation via sampling \\ $
         \hat{\Xb} =  {(\mathbf{1}_{p\times n} - \Mb) \:  \odot \Zb}  \; +  {\Mb \odot \Xb}$ \end{tabular} & Mean, model-based, multiple imputations & \eqref{eq:general_imputation}  \\ 
         \cdashline{2-4}[.4pt/2pt] 
         & \begin{tabular}[c]{@{}c@{}} Imputation as an optimization problem \\  $\hat{\Xb} = \argminn_{\Xb} \; d(\Xb,\Mb) + \alpha \, g(\Xb; {\thev}) + \lambda \, h(\Xb)$ \end{tabular} & Graph signal recovery, matrix completion & \eqref{eq:X_imputation}  \\ \hline
         \multirow{6}{*}{Estimation} & \begin{tabular}[c]{@{}c@{}}  Ignorable missing-data mechanism \\ $\hat{\thev} = \argmaxx_{\boldsymbol{\thev}} f(\Xbo;\thev)$ \end{tabular} & EM and its variants & Table \ref{Stochastic_EM_variants} \\ \cdashline{2-4}[.4pt/2pt] 
          & \begin{tabular}[c]{@{}c@{}}  Non-ignorable missing-data mechanism \\ $\hat{\thev} = \argmaxx_{\boldsymbol{\theta}} f(\Xbo,\Mb;\thev)$ \end{tabular} & Stochastic EM and its variants & Table \ref{Stochastic_EM_variants} \\
          \cdashline{2-4}[.4pt/2pt] 
           &  \multirow{2}{*}{ \begin{tabular}[c]{@{}c@{}}Prior  information \\ $\hat{\boldsymbol{\theta}}= \argminn_{\boldsymbol{\theta} 
           } d(\boldsymbol{\theta},\Xbo) + \lambda R(\boldsymbol{\theta})$ \end{tabular}} & A priori: MAP-EM & \eqref{prior_information} \\
           \cdashline{3-4}[.4pt/2pt] 
           & & Reparametrization based & \eqref{eq:m_step_structured} \\
           \hline
           \multirow{5}{*}{Learning} & \multirow{3}{*}{ \begin{tabular}[c]{@{}c@{}} Generative approaches for imputation \\
              $\hat{\Xb}_{\textsf{m}} = {I}_{\hat{\boldsymbol{\zeta}}} (\Xbo,\Mb),  
   \hat{\boldsymbol{\zeta}} = \argminn_{\boldsymbol{\zeta}}
   \mathbb{E}
   \{  \mathcal{D}_{\boldsymbol{\zeta}}(\Xbo, \Mb) \}$ \end{tabular}} & MVAE, MIWAE: KL divergence & \eqref{elbo_mvae_missing} \\ \cdashline{3-4}[.4pt/2pt] 
            & & GAIN: adversarial loss & \eqref{eq:GAIN} \\ \cdashline{3-4}[.4pt/2pt] 
            & & Optimal transport: Wasserstein distance & \eqref{eq:opti}  \\ \cdashline{2-4}[.4pt/2pt] 
            & \multirow{3}{*}{ \begin{tabular}[c]{@{}c@{}} Prediction despite missing values \\
            Only $\Xtrain$ missing: estimate $f(\yv|\Xb)$ \\ 
             Both $\Xtrain$ and $\xnew$ missing: estimate $f(\yv|\Xbo)$ \end{tabular}}  &  &  \\
            & & Naive imputation then  debiased algorithms & \eqref{eq:gk_withNA} \\
            \cdashline{3-4}[.4pt/2pt]
            & & MIA method, Impute-then-regress strategies & Figure \ref{fig:predict_both}

    \end{tabular}
    \caption{\textcolor{black}{Section-by-section overview of the tutorial, highlighting the unified formulations, key covered methods, and the corresponding references in the text.}}
    \label{tab:overview_general}
\end{table}

Third, we reinterpret key statistical concepts through a consistent formalism familiar to the SP community, making foundational theory accessible while connecting it to modern algorithmic implementations, including deep learning approaches.

We begin by introducing the adopted notations through the tutorial, followed by a formal definition of key concepts to model missing data. The tutorial then proceeds through three main sections corresponding to the three fundamental tasks: imputation, estimation, and learning. Finally, we conclude with a general takeaway and offer perspectives on open challenges and future directions for research. Table \ref{tab:overview_general} provides a useful guideline for readers interested in a general overview of the material covered in this tutorial.

\section*{Basic concepts of missing data}
\label{sec:basic}

This section defines fundamental concepts for handling missing data: the missing-data pattern, \textcolor{black}{the missing-data mechanism, and the notion of ignorability.} We represent the data by a $p \times n$ matrix $\Xb = [\xv_1 \; \dots \; \xv_n]$, where variables (or features) are arranged along the rows, and iid observations (or samples) $\xv_j$, for $j = 1, \dots, n$, are arranged along the columns. A summary of the notation used throughout this tutorial is provided in Table~\ref{tab:notation}. 

\begin{table}[t]
    \centering
    \begin{tabular}{ll|ll}
        $p$ & number of variables or features &
        $n$ & number of observations or samples \\
        $\bm{X}$ / $X_{i,j}$ & $p \times n$ data matrix / its $(i,j)$-th entry & 
        $[\Xb]_{i,:}$ & the $i$-th row of $\Xb$ \\
        $\bm{M}$ & missing-data pattern & $\bm{x}_j$ / $\bm{m}_j$ & the $j$-th sample/mask (column of $\Xb$/$\Mb$) \\
        $\bm{X}_{\sf m}$ / $\bm{X}_{\sf o}$ & missing / observed part of the data &
        $[\xv]_i$ & the $i$-th element of $\xv$ \\
        $\xv_{\textsf{m},j}$ / $\xv_{\textsf{o},j}$ & missing / observed part of sample $\xv_j$ &
        $\bm{\Sigma}$ & covariance or shape matrix  \\ 
        $\normop{\Xb}_{1,1}$ / $\normop{\Xb}_{1,\text{off}}$ & \lone~norm of $\Xb$ / off-diagonal entries of $\Xb$ &
        $\bm{\mu}$ & mean or location vector  \\
        $\bm{\theta}$ & vector of unknown parameters & $\odot$ & element-wise (Hadamard) product \\
        $Q(\thev;\thev^{(k)})$ & surrogate function at step $k$ of EM & $f(\cdot)$ & pdf or pmf \\
        $\mathbb{E}_{\xv \sim f(\cdot)}[g(\xv)]$ & expectation of $g(\xv)$ with $\xv \sim f(\cdot)$ &
        $\text{tr}(\cdot)$ / $(\cdot)^\T$  & trace operator / transpose operator\\
        ${\det}(\cdot) /{\det}^\ast(\cdot)$ & determinant / pseudo-determinant operator &
       {\color{black} $f(\cdot;\thev)$}  & {\color{black} Identical to $f(\cdot\vert \thev)$ when $\thev$ is deterministic}  \\
        $\overset{\text{EVD}}{=}$ / $\overset{\text{SVD}_\lambda}{=}$ & eigenvalue / soft thresholded  SVD &
        $\succeq \zerov$ & is positive semi-definite
    \end{tabular}
    \caption{Main notation used throughout the tutorial.}
    \label{tab:notation}
\end{table}

\subsection*{Missing-data pattern}
\label{sec:pattern}

To handle missing data systematically, it is convenient to define the missing-data \textit{pattern}~\cite{little2019}, which characterizes the structure and shape of missingness. We use an indicator matrix $\Mb \in \{0,1\}^{p \times n}$ such that, for the $i$-th element of the $j$-th sample of the data matrix $\Xb$,
\begin{equation}
\label{eq:mask}
   M_{i,j} = \left\{
    \begin{array}{ll}
        0 & \mbox{if } X_{i,j} \mbox{ is missing}, \\
        1 & \mbox{if } X_{i,j} \mbox{ is observed.}
    \end{array}
\right. 
\end{equation}
Figure~\ref{fig:Xo_Xm} illustrates this notation, showing how the complete data matrix $\Xb$ is partitioned into observed ($\Xbo$) and missing ($\Xbm$) components. \textcolor{black}{Note that the data we actually observe is $\Yb=\Mb \odot \Xb$}.

\begin{figure}
\centering
\resizebox{0.6\textwidth}{0.29\textwidth}{%
\begin{tikzpicture}[>=stealth]

\node at (0,1.7) {$\Xb$ (not fully observed)};

\matrix (X) [matrix of nodes,
             nodes={draw, minimum width=0.9cm, minimum height=0.9cm, text width=0.5cm, anchor=center, font=\small}]
{
    $X_{1,1}$ & $X_{1,2}$ & $X_{1,3}$ & $X_{1,4}$ & $X_{1,5}$ \\
    $X_{2,1}$ & $X_{2,2}$ & $X_{2,3}$ & $X_{2,4}$ & $X_{2,5}$ \\
    $X_{3,1}$ & $X_{3,2}$ & $X_{3,3}$ & $X_{3,4}$ & $X_{3,5}$ \\
};

\node at (0,-2.3) {$\Mb$ (observed)};

\matrix (M) [matrix of nodes,
             nodes={draw, minimum width=0.9cm, minimum height=0.9cm, text width=0.5cm,  anchor=center, font=\small},
             below=1cm of X]
{
    1 & |[fill=gray!30]| 0 & 1 & 1 &  |[fill=gray!30]| 0 \\
    |[fill=gray!30]| 0 & 1 & |[fill=gray!30]| 0 & |[fill=gray!30]| 0 & 1 \\
    |[fill=gray!30]| 0 & |[fill=gray!30]| 0 & |[fill=gray!30]| 0 & 1 & 1 \\
};

\node at (6,-.3) {$\Yb = \Mb \odot \Xb $ (observed)};

\matrix (XM) [matrix of nodes,
              nodes={draw, minimum width=0.9cm, minimum height=0.9cm, text width=0.5cm, font=\small, anchor=center},
              anchor=center,
              at={($(X)!0.5!(M) + (6,0)$)}]
{
    $X_{1,1}$ & |[fill=gray!30]| 0 & $X_{1,3}$ & $X_{1,4}$ & |[fill=gray!30]| 0 \\
    |[fill=gray!30]| 0 & $X_{2,2}$ & |[fill=gray!30]| 0 & |[fill=gray!30]| 0 & $X_{2,5}$ \\
    |[fill=gray!30]| 0 & |[fill=gray!30]| 0 & |[fill=gray!30]| 0 & $X_{3,4}$ & $X_{3,5}$ \\
};

\node at (12.2,1.7) {$\Xbo$ (observed)};
\matrix (Xo) [matrix of nodes,
              nodes={draw, minimum width=0.9cm, minimum height=0.9cm, text width=0.5cm, font=\small, anchor=center},right= 7cm of X]
{
    $X_{1,1}$ & |[fill=gray!30]| & $X_{1,3}$ & $X_{1,4}$ & |[fill=gray!30]| \\
    |[fill=gray!30]| & $X_{2,2}$ & |[fill=gray!30]| & |[fill=gray!30]| & $X_{2,5}$ \\
    |[fill=gray!30]| & |[fill=gray!30]| & |[fill=gray!30]| & $X_{3,4}$ & $X_{3,5}$ \\
};

\node at (12.2,-2.3) {$\Xbm$ (not observed)};
\matrix (Xm) [matrix of nodes,
              nodes={draw, minimum width=0.9cm, minimum height=0.9cm, text width=0.5cm, font=\small, anchor=center}, 
              below= 1cm of Xo]
{
    |[fill=gray!0]| & |[fill=gray!30]| $X_{1,2}$ &  |[fill=gray!0]| & |[fill=gray!0]| & |[fill=gray!30]| $X_{1,5}$ \\
     |[fill=gray!30]| $X_{2,1}$ & |[fill=gray!0]| &|[fill=gray!30]|  $X_{2,3}$ & |[fill=gray!30]| $X_{2,4}$ & |[fill=gray!0]|  \\
    |[fill=gray!30]| $X_{3,1}$ & |[fill=gray!30]| $X_{3,2}$ & |[fill=gray!30]| $X_{3,3}$ & |[fill=gray!0]|  & |[fill=gray!0]|  \\
};

\draw[->, thick] (X.east) -- (XM.west);
\draw[->, thick] (M.east) -- (XM.west);

\draw[->, thick] (XM.east) -- (Xo.west);
\draw[->, thick] (XM.east) -- (Xm.west);

\end{tikzpicture}
}
\caption{\textcolor{black}{Illustration of the missing-data pattern. White cells indicate observed entries and gray cells indicate missing entries. The observed matrix is $\Yb=\Mb \odot \Xb$, where the zero cases correspond to missing entries. The complete data matrix $\Xb$ is decomposed into observed ($\Xbo$) and missing ($\Xbm$) components using the indicator matrix $\Mb$.}}
\label{fig:Xo_Xm}
\end{figure}
\begin{figure}
	\begin{minipage}{.245\linewidth}
		\centering
		\subfloat[Multivariate]{\label{fig:pattern:multi}\includegraphics[scale=.6]{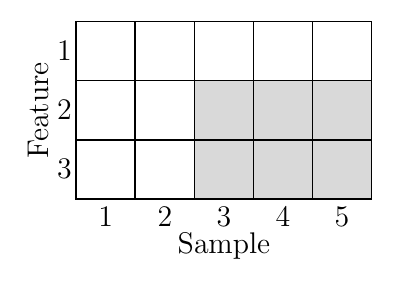}}
	\end{minipage}%
	\begin{minipage}{.245\linewidth}
		\centering
		\subfloat[Monotone]{\label{fig:pattern:mono}\includegraphics[scale=.6]{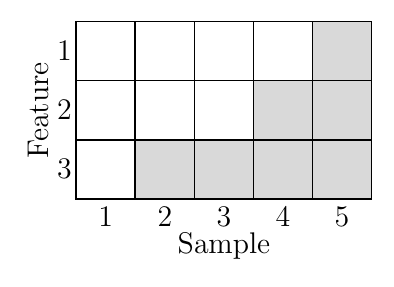}}
	\end{minipage}
	\begin{minipage}{.245\linewidth}
		\centering
		\subfloat[File-matching]{\label{fig:pattern:file}\includegraphics[scale=.6]{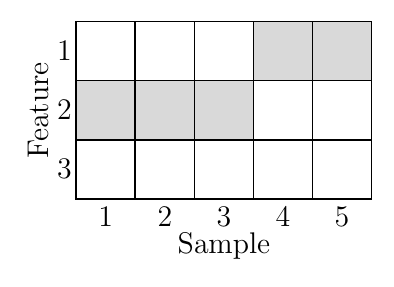}}
	\end{minipage}
	\begin{minipage}{.245\linewidth}
		\centering
		\subfloat[General]{\label{fig:pattern:gen}\includegraphics[scale=.6]{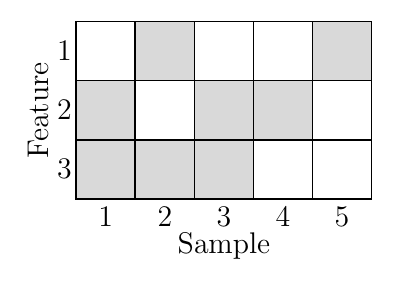}}
	\end{minipage}%
	\caption{\textcolor{black}{Different types of missing-data patterns for multivariate data with $p=3$ features and $n=5$ samples. White cells indicate $\Xbo$; gray cells indicate $\Xbm$.}}
	\label{fig:pattern}
\end{figure}

Missing-data patterns can be classified into five \textcolor{black}{non mutually exclusive} categories, some of which are illustrated in Figure~\ref{fig:pattern}:
\begin{itemize}
    \item[(a)] \textit{Univariate} missing data occur when a single variable is missing for some observations. For example, in environmental or industrial monitoring, a single sensor (e.g., humidity) may fail to report readings at certain time instances.
    
    \item[(b)] \textit{Multivariate} missing data arise when multiple variables are missing simultaneously. For example, wearable health devices sensing multiple physiological signals (e.g., heart rate, skin temperature, blood pressure) may fail together due to transmission loss.
    
    \item[(c)] The \textit{monotone} pattern occurs when the incomplete data can be sorted such that the $i$-th variable is at least as observed as the $(i+1)$-th variable for $i = 1, \dots, p-1$. A common example arises in sensor arrays, where progressive sensor failures systematically lead to increasing missingness.
    
    \item[(d)] The \textit{file-matching} pattern occurs in multimodal contexts where at least two sets of variables are never jointly observed. This is often seen when two sensing systems observe the same phenomenon but operate at different sampling frequencies.
    
    \item[(e)] The \textit{general} pattern indicates that multiple blocks of data are missing across multiple variables in a non-structured way. This is typical in imaging systems where pixels or regions are missing due to occlusions across multiple images.
\end{itemize}

\subsection*{Missing-data mechanism}
\label{sec:mechanism}

\textcolor{black}{The \textit{cause} of missingness, not just its pattern, profoundly influences the choice of appropriate statistical inference techniques. Before Rubin's seminal work~\cite{rubin1976}, the missing-data pattern $\Mb$ was \textcolor{black}{mostly} treated as a deterministic quantity and largely ignored in statistical analysis.} Treating $\Mb$ as a random variable fundamentally changed the theory and practice of working with incomplete data, providing a principled framework for dealing with uncertainty in the missingness itself.

\textcolor{black}{The missing-data pattern describes \textit{which} values are missing, but not the \textit{relationship} between missingness and data values. This relationship is characterized by the \textit{missing-data mechanism}, formally defined as follows.}

\begin{definition}[Missing-data mechanism]
\label{def:mechanism}
Assume that the missing-data pattern $\Mb \sim f(\Mb | \Xb; \phiv)$, where $f(\cdot)$ denotes the probability mass function of $\Mb$ parameterized by $\phiv \in \Omega_{\phiv}$. Three scenarios can be distinguished:
\begin{itemize}
    \item \textit{Missing completely at random (MCAR)}: The probability of a value being missing is independent of the data values, i.e., $f(\Mb | \Xb; \phiv) = f(\Mb; \phiv)$.
    
    \item \textit{Missing at random (MAR)}: The probability of missingness depends only on the observed values, i.e., $f(\Mb | \Xb; \phiv) = f(\Mb | \Xbo; \phiv)$.
    
    \item \textit{Missing not at random (MNAR)}: The probability of missingness depends on the values of the missing variables themselves, and possibly on the observed ones.
\end{itemize}
\end{definition}
\textcolor{black}{The definitions of missing-data mechanisms are frequently subject to discussion \cite{seaman2013meant}. In this tutorial, we assume that the above statements apply to all possible data patterns and values, corresponding to the everywhere mechanisms described by \cite{seaman2013meant}. Some concrete examples of these mechanisms are presented in the box \textit{Modeling Missing Measurements in Wearable Health Signals}.}

\subsection*{Ignorability and likelihood inference}
\label{sec:ignorability}

The distinction among MCAR, MAR and MNAR is critical because it determines which methods yield valid inference, as explained below. Under MCAR or MAR, many standard techniques remain applicable; under MNAR, failing to model the missingness mechanism can produce severely biased estimates.

In most applications, the parameter $\phiv$ governing the missing-data mechanism is not of primary interest. Statistical inference typically targets the parameter $\thev \in \Omega_{\thev}$ of the data distribution. \textcolor{black}{This leads to the notion of \textit{ignorability}, a key concept that determines whether the missing-data mechanism must be explicitly modeled for valid inference.}

\begin{definition}[Ignorable missing-data mechanism]
\label{def:ignorable}
A missing-data mechanism is \textit{ignorable} for likelihood inference if the maximum likelihood estimate of $\thev$ obtained from the observed-data likelihood $f(\Xbo; \thev)$ coincides with that obtained from the joint likelihood $f(\Xbo, \Mb; \thev, \phiv)$.
\end{definition}

Under MCAR or MAR, \textcolor{black}{if $\phiv$ is a nuisance parameter,} the maximum likelihood estimation (MLE)
$$\thev_{\rm MLE}, \phiv_{\rm MLE} 
    = \argmaxx_{(\thev,\phiv) \in \Omega_{\thev,\phiv}} \log f(\Xbo, \Mb; \thev, \phiv), $$
simplifies as follows
\textcolor{black}{$$\thev_{\rm MLE}
    = \argmaxx_{(\thev) \in \Omega_{\thev}} \log f(\Xbo; \thev). $$}
\textcolor{black}{Developing the joint log-likelihood leads to:}
\begin{align}
    \notag
    \log f(\Xbo, \Mb; \thev, \phiv)&= \log \int_{\mathcal{X}_{\sf m}} f(\Xb; \thev) \, f(\Mb | \Xb; \phiv) \, d\Xbm \\
    \notag
    &\stackrel{\text{MCAR/MAR}}{=}  \log \left[ \int_{\mathcal{X}_{\sf m}} f(\Xb; \thev) \, d\Xbm \cdot f(\Mb | \Xbo; \phiv) \right] \\
    \label{eq:ignorability}
    &\stackrel{(\thev, \phiv) \text{ distinct}}{=} \log f(\Xbo; \thev) +  \log f(\Mb | \Xbo; \phiv),
\end{align}
where $f(\Xbo; \thev) = \int_{\mathcal{X}_{\sf m}} f(\Xb; \thev) \, d\Xbm$ denotes the \textit{observed-data likelihood}.

\begin{lemma}[Ignorability for likelihood inference \cite{little2019}]
\label{lem:ignorability}
The missing-data mechanism is ignorable for likelihood inference if:
\begin{enumerate}
    \item The mechanism is MCAR or MAR, and
    \item The parameters $(\thev, \phiv)$ are such that 
    $\Omega_{\thev,\phiv} = \Omega_{\thev} \times \Omega_{\phiv}$. 
\end{enumerate}
\end{lemma}

When ignorability holds, likelihood-based inference can proceed using only the observed-data likelihood $f(\Xbo; \thev)$, without explicitly modeling the missingness mechanism, which is a substantial simplification in practice.

\begin{tcolorbox}[title=Application 1 : Modeling Missing Measurements in Wearable Health Signals, 
    title filled=false, 
    colback=gray!5!white, 
    colframe=gray!55!black, 
    sidebyside, 
    sidebyside align=top, 
    lower separated=false, 
    breakable, 
    bicolor,
    before skip=0.5\baselineskip,
    after skip=0pt]

To illustrate the three missing-data mechanisms, consider a health monitoring system that records two variables over time: heart rate (measured via a wearable device) and blood pressure (measured via a cuff-based monitor). Missing data can arise for different reasons, each corresponding to a distinct mechanism.

\textit{(i) MCAR scenario:} The wearable device occasionally experiences technical failures, such as battery depletion or signal dropout. These failures occur independently of the actual measurements, so the remaining data are representative of the complete data (see Figure~\ref{fig:MCAR}). Simple approaches such as mean imputation or ignoring missing entries may suffice.

\textit{(ii) MAR scenario:} Older patients may be less likely to complete blood pressure measurements regularly due to age-related forgetfulness or unfamiliarity with the technology. Here, missingness depends on an observed variable (age), making the mechanism ignorable (see Figure~\ref{fig:MAR}). More sophisticated algorithms that exploit the relationship between observed variables can effectively handle this case.

\textit{(iii) MNAR scenario:} Patients may avoid checking their blood pressure when they suspect it is high, perhaps due to anxiety or denial. Missingness now depends on the unobserved value itself, so the remaining data systematically underrepresent high blood pressure readings (see Figure~\ref{fig:MNAR}). Ignoring this mechanism leads to biased estimates; here, a valid inference requires explicitly modeling the relationship between missingness and the missing values.

\tcblower

\begin{center}
\begin{minipage}[t]{0.99\linewidth}
\centering
    \vspace*{0pt}
    \includegraphics[scale=0.35]{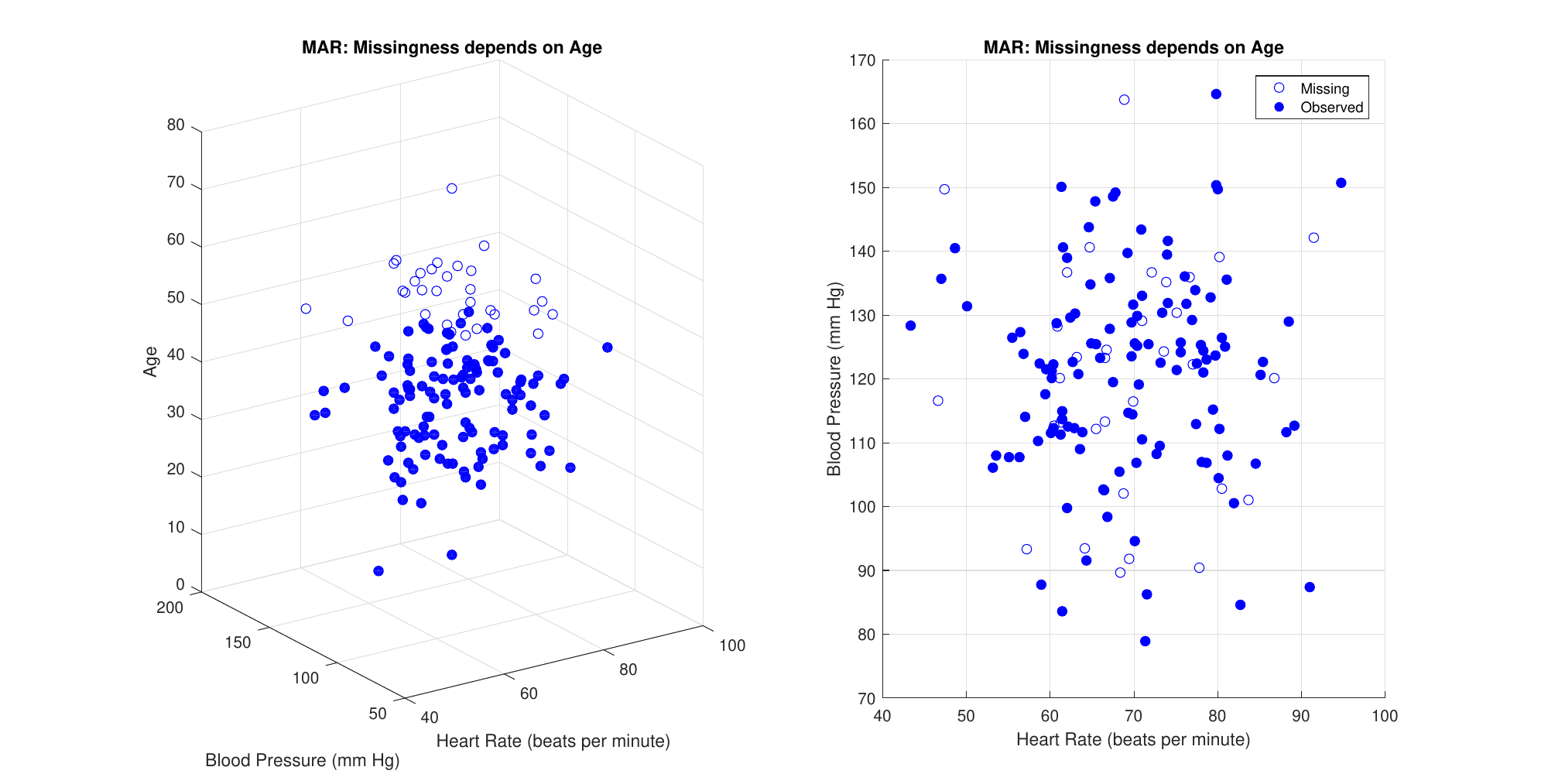}
    \captionof{figure}{MAR mechanism. Missing data depends on observed data: when adding variable age (left), we notice that patients older than 50 years may be less likely to wear the monitoring device consistently.}
    \label{fig:MAR}
\end{minipage}
\end{center}

\begin{center}
\begin{minipage}[t]{0.43\linewidth}
\centering
    \vspace*{0pt}
    \includegraphics[scale=0.35]{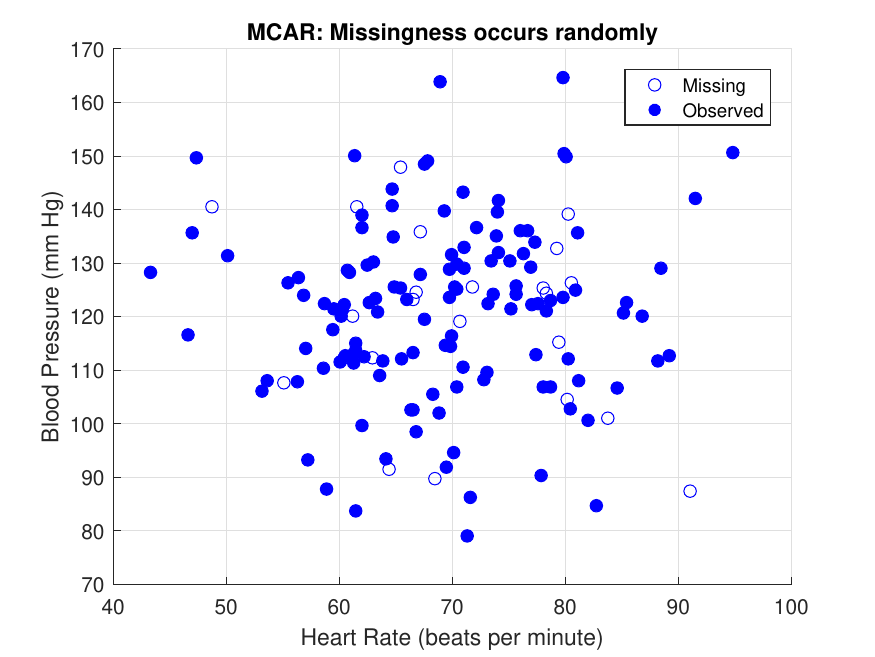}
    \captionof{figure}{MCAR mechanism. Missing data are independent of the data values.}
    \label{fig:MCAR}
\end{minipage}
\hfill
\begin{minipage}[t]{0.43\linewidth}
\centering
    \vspace*{0pt}
    \includegraphics[scale=0.35]{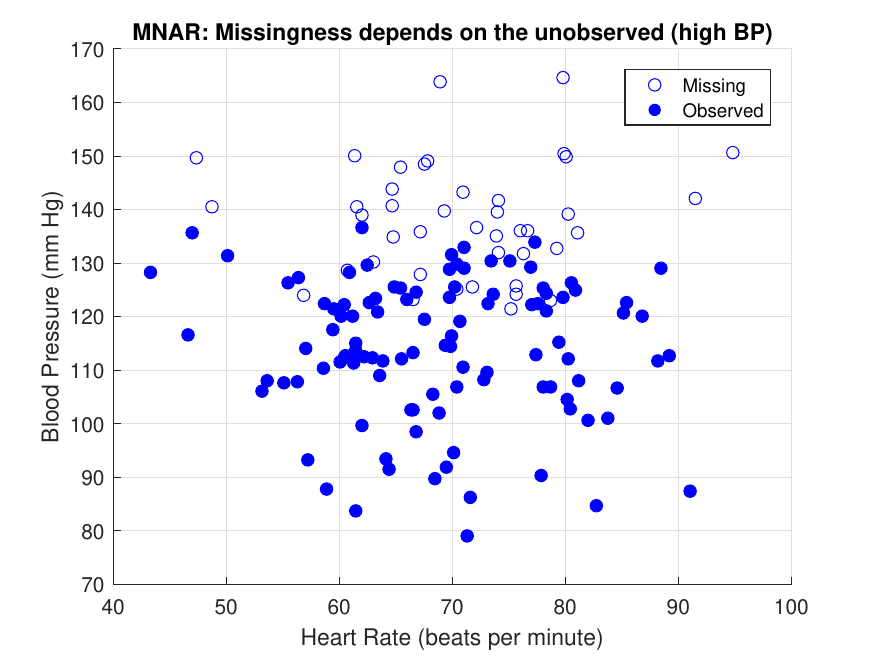}
    \captionof{figure}{MNAR mechanism. Missing data depend on unobserved data.}
    \label{fig:MNAR}
\end{minipage}
\end{center}

\end{tcolorbox}

\section*{Missing data imputation} 
\label{sec:imp}

{\color{black} 
Now equipped with the basic tools to deal with missing values, let us first explore the task of missing data imputation. In this section, we discuss classical and more recent imputation approaches for different types of signals, including graphs and low-rank matrices. The box \textit{Multiple Imputation of Financial Time Series} also shows an application to time series data. 

Imputation yields a complete dataset that can be used directly with standard analysis techniques requiring fully observed data, such as parameter estimation, classification, or regression. Additionally, imputation can serve as an initialization step for more sophisticated methods. At this point, one might already wander: Why the need of imputing ? Why not simply removing samples containing any missing values ? This approach, known as \textit{list-wise deletion} or \textit{complete-case analysis}, is only justifiable when the remaining data is representative of the original distribution and when the proportion of missing values is small. In many applications, deletion is problematic: for example, discarding all segments of an EEG containing transient artifacts may eliminate critical event-related information, fundamentally altering the interpretation of neural activity. 

In the following, an imputed dataset, denoted $\hat{\Xb}$, is the result of the following operation:
\begin{equation}\label{eq:general_imputation}
\hat{\Xb} =  \underbrace{(\mathbf{1}_{p\times n} - \Mb) \:  \odot \Zb}_{\text{imputed missing values}} \quad \; +  \underbrace{\Mb \: \odot \Xb}_{\text{unchanged observed values}}  ,
\end{equation}
where $\mathbf{1}_{p\times n}$ is a matrix of size $(p,n)$ with all entries equal to one and $\Zb$ is a matrix containing values that replace the values of $\Xbm$.} 

\subsection*{Classical imputation methods}
The simplest approach is probably \textit{mean imputation}: replace each missing entry with the mean of observed values for that variable, that is \textcolor{black}{$Z_{i,j} = (\sum_j M_{i,j}X_{i,j})/\sum_j M_{i,j}, \: \forall i\in \{1,\dots,p\}, \forall j\in \{1,\dots,n\}$. While easy to implement, this method ignores relationships between variables and artificially reduces variance. A modest improvement is to impute using the conditional mean given observed values in the same sample, plus a random residual to preserve empirical variance. A classic yet efficient alternative is to predict missing values from similar samples in the dataset~\cite{troyanskaya2001missing}. For each sample with missing data, one identifies its $k$ nearest neighbors (using a distance metric on complete variables) and imputes by averaging their values. This method preserves local structure in the data but faces practical challenges: (1) it requires defining an appropriate distance metric over incomplete variables, (2) the distance computation becomes unreliable when many variables are missing, and (3) selecting $k$ is non-trivial and data-dependent. In general, simple methods suffer clear limitations: they fail to capture complex dependence structures and perform poorly when missingness is substantial. However, they serve two important roles: as interpretable baselines for benchmarking more sophisticated techniques, and as practical solutions when missing data are truly sparse (e.g., $<5\%$) and missing completely at random (MCAR). In the following, we present more sophisticated imputation methods, dedicated for MCAR or MAR values (e.g., when the missing-data mechanism is ignored, see \eqref{eq:ignorability}).} 


\subsection*{Model-based imputation methods}
\textcolor{black}{
In \textit{model-based} imputation methods, the entries of $\Zb$ are sampled from the conditional distribution of the missing part of the data given the observed one, i.e.,
$\Zb \sim f(\Xbm|\Xbo)$\footnote{There is a slight abuse of notation here. Rigorously, this denotes $\Zb \sim f(\cdot|\Xbo)$.}.
These approaches fall into two broad categories~\cite[Sections 4.4–4.6]{vanbuuren2018}: the \textit{iterative conditional} methods that directly learn 
$f(\Xbm|\Xbo)$ and  the \textit{full generative} methods that learn the joint distribution $f(\Xb) = f(\Xbo, \Xbm)$.
}
\subsubsection*{Iterative conditional methods} \textcolor{black}{This first strategy consists in directly estimating the conditional distribution $f(\Xbm \mid \Xbo)$ without specifying the full joint distribution $f(\Xb)$. This simple algorithm iterates until convergence:
\begin{enumerate}
    \item[(i)] For each variable $i$, fit a predictive model  {using complete samples of $[\Xb]_{i,:}$ (containing no missing values)},
     with $[\Xb]_{i,:}$ as the target and $[\Xb]_{-i,:}$ (other variables) as predictors.
    \item[(ii)] Use the fitted model to impute missing values in $[\Xb]_{i,:}$.
    \item[(iii)] Repeat for all variables.
\end{enumerate}
The predictor can take various forms, such as a random forest~\cite{stekhoven2012missforest}, linear regression, or other flexible models. A popular approach is multivariate imputation by chained equations (MICE)~\cite{vanbuuren2018}. While these methods adapt well to different data types, they are computationally expensive: each variable requires training a separate model, involving the estimation of $p$ different conditional distributions. Additionally, their convergence and theoretical properties rely on regularity conditions that may not always hold.}

\subsubsection*{Full generative methods}\textcolor{black}{ Alternatively, one can specify a parametric distribution for both complete data $\Xb$ and samples from the posterior conditional $f(\Xbm \mid \Xbo)$. For example, assuming that each observation $\xv_j \sim \mathcal{N}(\muv, \Sigmab)$ for $j = 1, \dots, n$, the parameters $\muv$ and $\Sigmab$ can be estimated via the Expectation-Maximization (EM) algorithm~\cite{honaker2011amelia} (detailed in the \textit{Estimation} section). 
{It is also the formulation adopted by several imputation methods discussed below, including matrix completion in the next subsection and generative approaches in the \textit{Learning} section.}}

\subsubsection*{Multiple imputation methods}
\textcolor{black}{In single imputation methods, the entries of $\Zb$ in \eqref{eq:general_imputation} are sampled only once. In contrast, \textit{multiple imputation} (MI) generates $M>1$ plausible imputations for each missing value, resulting in $M$ complete datasets.} Each dataset is analyzed independently, and the results are pooled using standard procedures \cite{rubin1976,little2019} to account for both within-imputation and between-imputation variability. This approach provides a more robust and reliable analysis compared to single imputation. A detailed example using financial time series is provided in the box at the end of this section.


\subsection*{Imputation as a constrained optimization problem}
\textcolor{black}{
In this section, we examine imputation approaches focusing on signals defined on graphs (\textit{graph signal recovery}) and signals with a low-rank structure (\textit{matrix completion}). Assume we have observations with missing entries as $\Yb = \Mb \odot \Xb$ (see Figure \ref{fig:Xo_Xm}). The imputed data matrix $\hat{\Xb}$ is obtained by solving a constrained optimization problem of the form\footnote{Note that Eq.~\eqref{eq:general_imputation} still applies as a general imputation framework for $\hat{\Xb}$. Due to the specific nature of the signals presented in this section (graphs and low-rank matrices), imputation largely benefits from the tools of optimization.} :
\begin{equation}
\label{eq:X_imputation}
    \hat{\Xb} = \argminn_{\Xb} \; d(\Xb,\Mb) + \alpha \, g(\Xb; {\color{black}\thev}) + \lambda \, h(\Xb),
\end{equation}
where $\alpha, \lambda \geq 0$ are regularization parameters. The three terms in this objective serve distinct purposes:
}
\begin{itemize} 
    \item \textcolor{black}{\textit{Fidelity term} $d(\Xb,\Mb)$: Measures the discrepancy between the reconstructed data and the available observations. A common choice is the squared Frobenius norm of the residual on observed entries.
    \begin{equation}\label{eq:fidelity_term}
    d(\Xb,\Mb) = \|\Mb \odot (\Yb - \Xb)\|_F^2.
    \end{equation}
    }
   \textcolor{black}{To handle outliers, one can also use the Huber-loss function for the fidelity term, defined as:
    \begin{align*}
        d(\Xb,\Mb) = \sum_i \sum_j H_\delta (M_{i,j}  (Y_{i,j}-X_{i,j})) \quad \text{with} \quad  H_\delta (x) =
    \frac{1}{2}x^2    \mathbbm{1}_{\vert x\vert\leq \delta} + 
    \delta \left(\vert x\vert - \frac{\delta}{2}\right)  \mathbbm{1}_{\vert x\vert > \delta},
    \end{align*}
     where $\delta$ is a constant (slope) parameter that approximately corresponds to the standard deviation of the inlier errors.}


    \item \textcolor{black}{\textit{Smoothness term} {\color{black}$g(\Xb;\thev)$: 
    Quantifies how smoothly the data conforms to the model structure encoded by the parameters $\thev$.
    For instance, in graph signal processing, this term enforces consistency with an underlying graph model, where $\thev$ represents the graph Laplacian or adjacency matrix.}}
    
    \item \textcolor{black}{\textit{Regularization term} $h(\Xb)$: Promotes desired properties of the solution, such as bounded energy (via the Frobenius norm, $h(\Xb) = \|\Xb\|_F^2$) or low-rank structure (via the nuclear norm, $h(\Xb) = \|\Xb\|_*$).}
\end{itemize}

\subsubsection*{Graph signal recovery}
\label{sec:graph_imputation}

{\color{black}
Many applications involve signals defined on graphs, such as sensor networks, social networks, or transportation systems. When such signals contain missing entries, imputation can leverage the underlying graph structure. 
A graph is typically represented by a (weighted) adjacency matrix $\Wb$, where the entry $W_{i,j}$ denotes the weight of the edge from node $i$ to node $j$.
}

\textcolor{black}{Using the framework in~\eqref{eq:X_imputation}, with the fidelity term from~\eqref{eq:fidelity_term} and $\thev = \Wb$, the {remaining} terms of \eqref{eq:X_imputation} can be distinguished as follows}:

{\color{black}
\textbf{Smoothness:} For an undirected graph with positively weighted edges, where $\Wb=\Wb^\T\geq \zerov$, the following definition of smoothness is commonly utilized \cite{perraudin2017}:
\begin{align}
\label{eq:graph_smooth}
    g(\Xb; \Wb) = \frac{1}{2} \sum_i \sum_j W_{i,j} \normop{[\Xb]_{i,:} - [\Xb]_{j,:}}_{p}^p. 
\end{align}
Here, $\normop{\cdot}_p$ denotes the \lp~vector norm, typically chosen as the \lone~ or \ltwo~norm. For $p=2$, the term $g(\Xb; \Wb)$ is quadratic in $\Xb$ and is also referred to as \textit{graph Tikhonov regularization}. Several works also employ a $p=1$ variant of the graph-based smoothness, known as \textit{graph total variation}, which is suitable for edge-preserving denoising. Based on this measure, a graph signal is  smooth when the signal values at connected nodes are similar, with the degree of similarity proportional to the weight of the connection    $W_{i,j}$. 
For graph signals with a temporal dimension (e.g., time series on a network), one can use \textit{spatio-temporal smoothness} by replacing $\Xb$ in \eqref{eq:graph_smooth} with the first-order temporal difference matrix $\Delta(\Xb) = [\xv_1, \xv_2 - \xv_1, \dots, \xv_n - \xv_{n-1}]$.
The resulting measure simultaneously enforces spatial smoothness (via graph-domain similarities) and temporal smoothness (via finite times differences).
For directed graphs, where $\Wb$ is non-symmetric and may have negative edge weights, a slightly different notion of smoothness can be defined by measuring signal variations across the edges as 
\begin{align}
    g(\Xb;\Wb) = \sum_{j=1}^n\normop{\xv_j -\frac{1}
    {\vert\lambda\vert_{\max}(\Wb)}
    \Wb\xv_j}_p^p,
\end{align}
where $\vert\lambda\vert_{\max}(\Wb)$ denotes the eigenvalue of $\Wb$ with the largest magnitude in absolute value \cite{chen2015}.
This formulation, with $p=1$ and $p=2$, gives alternative definitions for graph total variation and  Tikhonov regularization for directed graphs, respectively. 

\textbf{Regularization:} 
In graph signal imputation, the regularization term is frequently selected as the Frobenius norm $h(\Xb) = \normop{\Xb}_F^2$, or simply omitted by setting $h(\Xb) = 0$.

Using the smoothness term   $g(\Xb;\Wb)$ with $p\geq 1$ and  a convex function  $h(\Xb)$, the  graph signal recovery problem becomes convex and can be efficiently solved using standard convex optimization  methods such as ADMM, proximal gradient or majorization minimization.
}

\subsubsection*{Low-rank matrix completion}
\label{sec:matrixcompletion}

\textit{Low-rank matrix completion} addresses the recovery of a matrix $\Xb$ from partial observations $\Yb = \Mb \odot \Xb$ under the assumption of low-rank structure. The ideal problem is:
\begin{equation}
\minn_{\Xb} \quad \text{rank}(\Xb) \quad \text{subject to} \quad \Mb \odot (\Xb - \Yb) = \mathbf{0}.
\end{equation}
This non-convex problem is intractable due to the non-convex rank constraint, which makes it in general NP-hard.
\textcolor{black}{A standard convex relaxation replaces the rank constraint with the nuclear norm, fitting the general framework in~\eqref{eq:X_imputation}, with fidelity term given in \eqref{eq:fidelity_term} and:
\begin{align}
    \text{\textbf{Smoothness:}} \quad &g(\Xb;\thev) = 0 \\
    \text{\textbf{Regularization:}} \quad &h(\Xb) = \|\Xb\|_*,
\end{align}}
where 
$\|\Xb\|_*$ denotes the nuclear norm  (sum of singular values). 
The absence of a smoothness term reflects the assumption that the data matrix itself, through its low-rank structure, contains sufficient organization. This formulation is widely used in recommendation systems, sensor array processing, and other applications where observations lie approximately in a low-dimensional subspace.


By noticing that $ \norm \Xb \norm_* = \minn\limits_{\Ub, \Vb: \Xb=\Ub\Vb^\T} 
\frac{1}{2}
\norm \Ub \norm^2_F + \frac{1}{2}\norm \Vb \norm^2_F $, where $\Ub \in \mathbb{R}^{p\times r}$, $\Vb \in \mathbb{R}^{n\times r}$ and $r$ is the rank of $\Xb$, the optimization problem in \eqref{eq:X_imputation} can be rewritten as:
\begin{equation}\label{eq:nuclear_norm_min}
    \minn_{\Ub, \Vb} \norm \Mb \odot (\Ub\Vb^\T-\Yb)\norm_F^2 + \lambda(  \norm \Ub \norm^2_F + \norm \Vb \norm^2_F ).
\end{equation}
This new formulation is very close to the problem of retrieving an underlying subspace $\Ub$ from the data $\Yb$ in the \textit{batch} setting, i.e., when $\Yb$ has a fixed number of observations (columns). It also shares clear connections with \textit{online} subspace estimation with missing data, also known as subspace tracking (ST), i.e., when the original data is gradually constructed from a data stream (treated in the \textit{Estimation} section).

To solve the optimization problem ~\eqref{eq:nuclear_norm_min}, classical methods rely on proximal iterative algorithms based on Singular Value Decomposition (SVD), where the solutions are usually computed using a grid-search on the regularization parameter $\lambda$. More specifically, a commonly used algorithm called softImpute \cite{hastie2015matrix} performs the following two steps iteratively, starting from an initial imputed data matrix $\hat{\Xb}^{(0)}$ (for example, using the mean imputation) and \textcolor{black}{a regularization parameter $\lambda$}:
\begin{enumerate}[label=(\roman*)]
   \item\label{matcomp_1} Compute the \textcolor{black}{soft thresholded SVD} of $\hat{\Xb}^{(k)}$, i.e.\ compute the SVD and \textcolor{black}{retain the singular values higher than a threshold $\lambda$, $\hat{\Xb}^{(k)}\overset{\text{SVD}_\lambda}{=} \Ub^{(k)}\Sb_\lambda^{(k)}\Vb^{(k)\T}$, with $\Sb_\lambda^{(k)}$ a diagonal matrix such that its diagonal terms are $[\Sb_\lambda^{(k)}]_{ii}=\max((\sigma_i-\lambda),0),$ with $\sigma_i$ the singular values of $\hat{\Xb}^{(k)}$, for $i\in \{1,\dots,p\}$.}
   \item \label{matcomp_2} Update the imputed data matrix (only on the missing values) with the quantity computed in Step \ref{matcomp_1}
   $\hat{\Xb}^{(k+1)}_{i,j} = (1-M_{i,j}){\left[\left(\Ub^{(k)}\Sb_\lambda^{(k)}\Vb^{(k)\T}\right)\right]_{i,j}}+M_{i,j}{X_{i,j}}$. 
\end{enumerate}

\begin{tcolorbox}[colback=gray!10!white, colframe=gray!50!black, 
    title=Application 2 : Multiple Imputation of Financial  Time Series, 
    sidebyside, sidebyside align=top,
    lower separated=false, 
    bicolor,
    breakable]

Financial data of traded assets frequently contain missing values due to non-trading periods, lack of liquidity, data collection gaps, or handling errors. Traditional interpolation methods include \textit{last observation carried forward} (LOCF), which repeats the most recent observed price, and \textit{linear interpolation} of prices (equivalent to mean interpolation of returns). While simple, these approaches fail to capture the stochastic nature of financial returns.

More principled imputation methods exploit structural knowledge about the data. In recommender systems, one leverages low-rank structure; in image inpainting, local geometric similarity. For financial time series, the key structure is \textit{temporal dependence}. One approach is to use the MICE method~\cite{vanbuuren2018} with reengineered lagged variables as inputs. Alternatively, one can directly specify a time series model, such as a random walk or autoregressive process, fit it to the observed data, and use Bayesian inference to impute missing values from the posterior distribution~\cite{liu2019parameter}. Figure~\ref{fig:interpolation_vs_stat_imputation} compares naive interpolation methods with MICE and the model-based ImputeFin approach, demonstrating the superior realism of statistical imputation. When multiple assets are modeled jointly, one can further exploit cross-sectional dependencies.

\begin{center}
\includegraphics[width=0.6\textwidth]{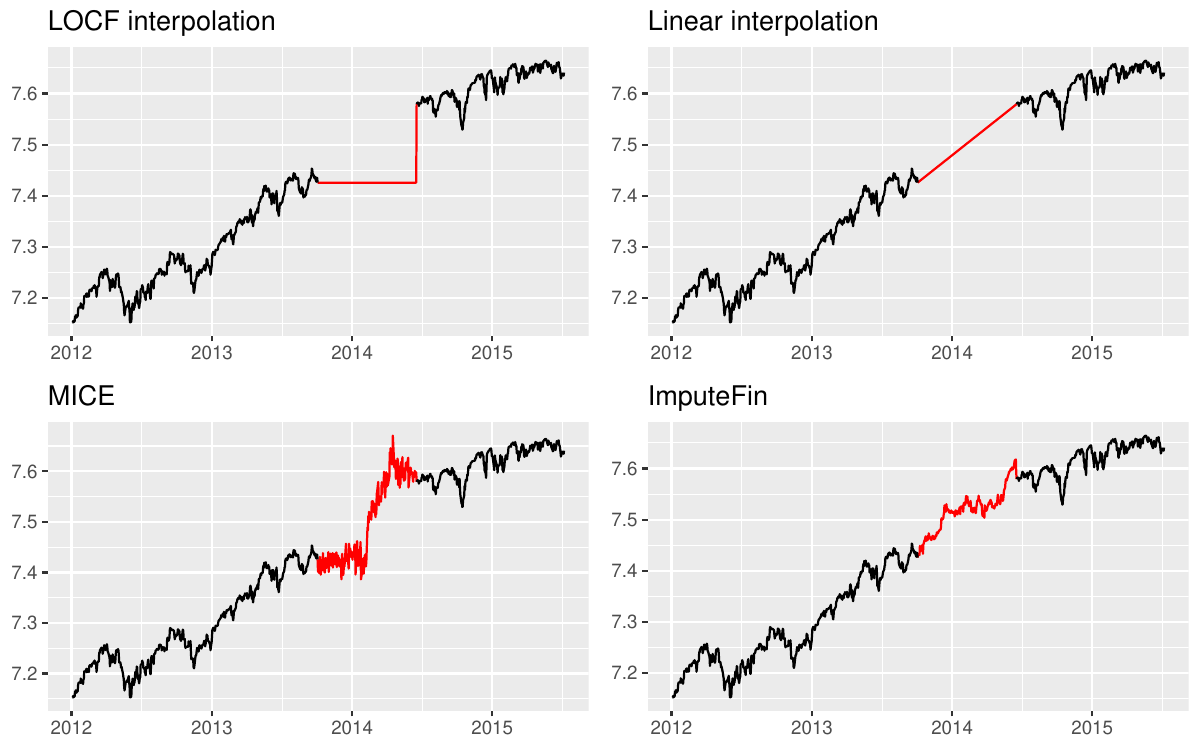}
\captionsetup{font=small}
\captionof{figure}{Comparison of naive interpolation, MICE with reengineered lagged variables, and model-based imputation (ImputeFin) for financial time series.}
\label{fig:interpolation_vs_stat_imputation}
\end{center}

In applications such as portfolio optimization, backtesting, and risk modeling, it is often desirable to generate multiple plausible imputations rather than a single point estimate, an approach known as \textit{multiple imputation}. A probabilistic model for the missing values naturally enables this: one simply draws from the estimated posterior distribution as many times as needed. Figure~\ref{fig:fin_multiple_imputation} shows {\color{black} the ground-truth S\&P~500 index   followed by three representative multiple imputations  obtained using the model-based approach of~\cite{liu2019parameter}, implemented in the R package \texttt{imputeFin}.\footnote{\url{https://CRAN.R-project.org/package=imputeFin}}}

\begin{center}
\includegraphics[width=0.6\textwidth]{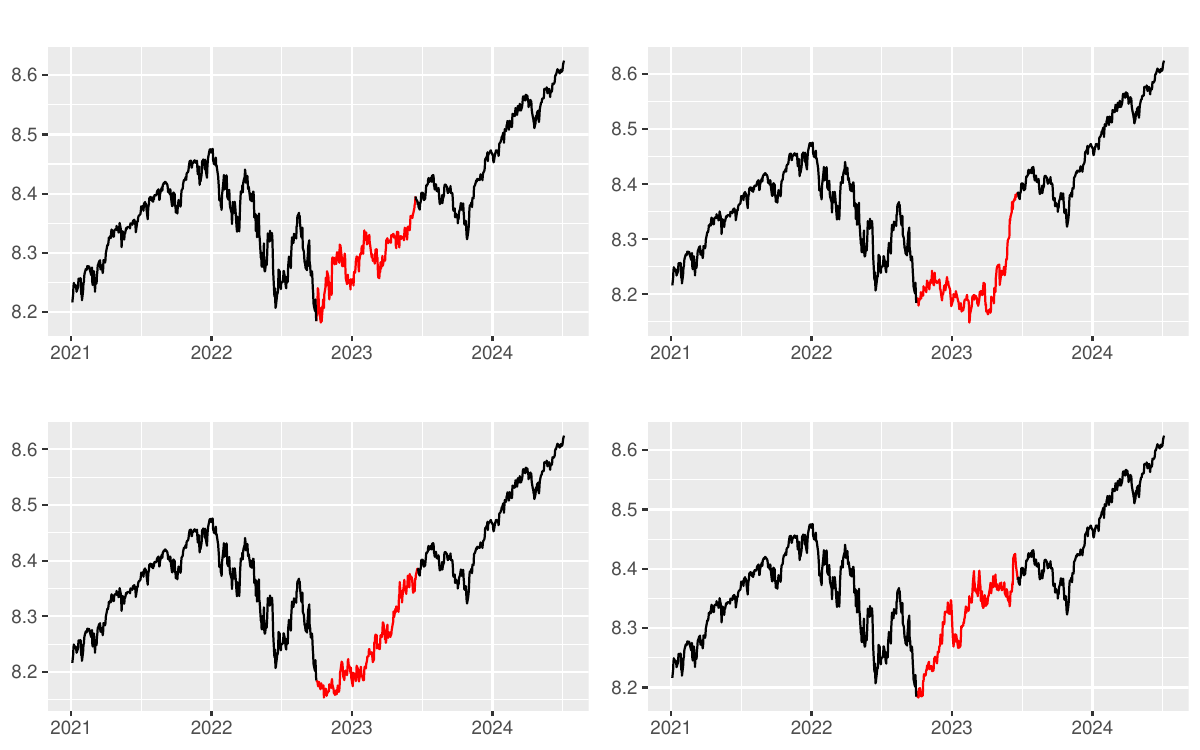}
\captionsetup{font=small}
\captionof{figure}{Multiple imputation of a financial time series. Can you distinguish the ground truth from the imputations?}
\label{fig:fin_multiple_imputation}
\end{center}

As the figures illustrate, model-based statistical imputation performs remarkably well.
The underlying method works as follows. To capture both temporal structure and the heavy-tailed nature of financial returns, an autoregressive (AR) model is assumed:
\begin{equation}
\label{eq:AR_orig}
x_t = \mu + a \, x_{t-1} + \epsilon_t,
\end{equation}
where $x_t$ is the time series value at time $t$, $\mu$ is the drift, $a$ is the AR coefficient, and the innovations $\epsilon_t$ follow a Student's $t$-distribution, $\epsilon_t \sim \mathrm{t}(0, \sigma^2, \nu)$, with scale $\sigma$ and degrees of freedom $\nu > 0$ controlling tail thickness. The parameter vector is $\thev = (a, \mu, \sigma, \nu)$.

The key challenge is estimating $\thev$ in the presence of missing values. With complete data and Gaussian innovations, this reduces to least squares; with Student's $t$ innovations, the problem becomes more involved but remains tractable. However, with missing values, maximum likelihood estimation requires solving
\begin{equation}
\hat{\thev}_{\text{MLE}} = \argmaxx_{\thev} \log f(\xbo ; \thev) = \argmaxx_{\thev} \log \int \prod_{t=2}^{n} f(x_t | x_{t-1}; \thev) \, d\xbm,
\end{equation}
where $\xbo$ and $\xbm$ denote the observed and missing components of the time series $\xv$. This integral lacks a closed-form solution due to the Student's $t$ likelihood. The method of~\cite{liu2019parameter} addresses this using the stochastic approximation EM (SAEM) algorithm  with multiple Monte Carlo samples, which will be discussed in the next section  (see Table~\ref{Stochastic_EM_variants}).

\tcblower
\end{tcolorbox}

{\color{black}

\paragraph*{Takeaway} For imputation, traditional methods such as mean imputation provide computationally efficient baselines but underestimate variability and may introduce bias, particularly {for MAR or} {MNAR data} and under complex dependency structures. 
Robust alternatives include model-based techniques which replace missing values with values sampled from the conditional distribution $f(\Xbm | \Xbo)$, such as multiple imputation (e.g., MICE). For data with specific structures, such as graph-structured signals and low-rank data structures, the imputed data can be expressed as the solution of a (constrained) optimization problem. Table~\ref{tab:gsp_vs_mc} provides a comparison between graph signal imputation and low-rank matrix completion methods.

}

\begin{table}[t]
\arrayrulecolor{black} 
\centering
\begin{tabular}{p{3cm}p{6.2cm}p{4.2cm}}
\hline
Aspect & Graph signal recovery & Low-rank matrix completion \\
\hline
Structural prior
& Graph smoothness 
& Low-rank structure \\
Optimization problem
& Often convex with \lone~(Total Variation),
& Originally non-convex \\
& \ltwo~(Tikhonov)  smoothness term & Convex with nuclear-norm
\\
Vectors vs.\ matrices
& Handles both vectors and  matrices 
& Primarily handles matrices \\
Graph dependence
& Requires a known or learned graph
& No graph in classical form \\
Typical algorithms
& Proximal/gradient methods, Alternating direction method of multipliers (ADMM) & {Soft-thresholded SVD}, ADDM \\
Complexity driver
& Size of the graph,
& Matrix size and rank, \\
& vertex domain operations (if used) & repeated SVDs \\
{Main} applications
& Sensor networks
& Image inpainting    
\\
\hline
\end{tabular}
\caption{\textcolor{black}{ Comparison between graph signal recovery and low-rank matrix completion.}}
\label{tab:gsp_vs_mc}
\end{table}

\section*{Estimation with Missing Data under a Parametric Model Framework } \label{sec:estim} 

In this section, we address estimation and inference in the presence of missing data, operating under a parametric model that reflects the underlying data-generating process. The primary methodological approach involves constructing a cost function derived from maximum likelihood estimation (MLE) estimation, under ignorable \textcolor{black}{and non-ignorable missing-data mechanisms. This approach can be further regularized to incorporate structural assumptions or prior knowledge available to the practitioner (such as low-rank constraints, parameter restrictions, or other domain-specific information.)}



\subsection*{Statistical estimation under ignorable mechanism}
\label{subsec:stat_estim}
Let us assume that the parametric model associated with a data-generating process and the missing mechanism for the $j$-th sample reads $f(\xv_j,\mv_{j};\thev,\phiv)$, in which $\thev$ represents the deterministic unknown parameter of interest (e.g., the mean $\bm{\mu}$ belonging to the set of  real or complex numbers or the data covariance matrix $\mathbf{\Sigma}$ belonging to
the  semidefinite positive matrices), $\phiv$ is the set of parameters governing the missing mechanism, $\xv_j$ and $\mv_j$ are the $j$-th sample and mask (see Table \ref{tab:notation}).
Under \textit{ignorable} missing-data mechanisms with multiple independent samples, the 
  MLE takes the form:
  $$  \hat{\thev}_{\rm MLE}
   = \argmaxx_{\thev \in \Omega_{\thev}}\sum_j \log \int  f({\xbo}_{,j},{\xbm}_{,j} ;\thev)   d {\xbm}_{,j}.$$
 Solving the latter in closed-form expression is generally infeasible.
An exception is the \textcolor{black}{MLE} of the mean and/or the covariance matrix of a multivariate normal distribution from incomplete data with a monotone pattern and under MCAR or MAR mechanism \cite{little2019,cherifi2025maximum}.
In most cases,
maximum likelihood estimates from incomplete data can be obtained using the Expectation-Maximization (EM)
algorithm or its variants. The EM algorithm is a natural choice for missing data problems because it considers missing values as latent variables. In  the E-step, the EM computes their conditional expectation, while the M-step updates parameters by maximizing this expectation. 
Specifically, at step $k=0$, we initialize the unknown parameter using a naive imputation method or one of the imputation approaches listed above, to obtain $\thev^{(0)}$. Then, the EM iterates between the following E- and M-steps until convergence:
\begin{itemize}
    \item \underline{\textit{E-step (Expectation)}}: It consists of computing a surrogate function denoted by $Q(\cdot)$, which
is the conditional expectation of the complete-data log-likelihood conditioned to
the observed data and the previous estimate of the parameter $\thev^{(k)}$. The complete-data are in general the collection of observed data and latent data,  {in our case $\Xb$ is a collection of $\Xbo$ and $\Xbm$, which leads to}
\begin{equation} \label{eq:e_step}
Q(\thev;\thev^{(k)})=  \mathbb{E}_{\Xbm\sim f(\Xbm|\Xbo; \thev^{(k)})}
[
\log f(\Xbo,\Xbm; \thev) ].
\end{equation} 
    \item \underline{\textit{M-step (Maximization)}}:  Updates the unknown parameter by maximizing the surrogate function
    \begin{equation} \label{eq:m_step}
    \thev^{(k+1)} =\argmaxx_{\thev \in \Omega_{\thev}} Q(\thev;\thev^{(k)}).
    \end{equation}
\end{itemize}
 
 The EM algorithm offers several advantages, including ease of implementation, a clear statistical interpretation of each step without ad-hoc imputation, and a general guarantee of convergence. However, a significant drawback is that a high proportion of missing information can result in a slow convergence rate. In the following, we briefly discuss different strategies to compute the EM for different data distributions.

\subsubsection*{The Gaussian distribution case}
\label{Gaussian_distribution_case}
Let us assume that the samples ${\xv}_j$, $j=1,\dots,n$, are independent and that  ${\xv}_j\sim \mathcal{N}(\muv,\Sigmab)$. Under MAR/MCAR, the surrogate function of the EM algorithm reads
$$
Q(\thev;\thev^{(k)}) \propto
 n \log \det(\Sigmab) -  \sum_j
 \mathbb{E}_{{\xbm}_{,j}\sim f({\xbm}_{,j}|{\xbo}_{,j}; \thev^{(k)})}
 [  (\xv_j-\muv)^\T \Sigmab^{-1} (\xv_j-\muv)].
$$
First, since $\xv_j$ contains the observed values ${\xbo}_{,j}$ and the missing ones ${\xbm}_{,j}$, the expectation is done only with respect to the missing values. Second, the term under the expectation contains the sufficient statistics  ${\xbm}_{,j}$ and
 ${\xbm}_{,j}{\xbm}_{,j}^\T$. Consequently, being able to compute $ \mathbb{E}_{{\xbm}_{,j}\sim f({\xbm}_{,j}|{\xbo}_{,j}; \thev^{(k)})}[{\xbm}_{,j}]$ and 
\linebreak
 $ \mathbb{E}_{{\xbm}_{,j}\sim f({\xbm}_{,j}|{\xbo}_{,j}; \thev^{(k)})}[{\xbm}_{,j}{\xbm}_{,j}^\T]$ gives us a closed-form expression of $Q(\thev;\thev^{(k)})$. In the Gaussian case, the aforementioned expectation can be easily derived by noticing that
${\xbm}_{,j}|{\xbo}_{,j}\sim \mathcal{N} (\muv_{\sf m|o},\Sigmab_{\sf m|o})$. Here, the rectified mean and the
covariance matrix read $$\muv_{\sf m|o}=\muv_{\sf m,m}-\Sigmab_{\sf m,o}\Sigmab_{\sf o,o}^{-1}({\xbo}_{,i}-\muv_{\sf o,o})$$ and $$ \Sigmab_{\sf m|o}=\Sigmab_{\sf m,m}-\Sigmab_{\sf m,o}\Sigmab_{\sf o,o}^{-1}\Sigmab_{\sf o,m}$$ in which
$\muv_{\sf m,m}$, $\Sigmab_{\sf m,m}$ and  $\Sigmab_{\sf m,o}$ denote respectively, the mean of the missing values, the covariance matrix of the missing values and the covariance between ${\xbm}$ and ${\xbo}$ \cite{little2019}.

\subsubsection*{The scaled Gaussian distribution case}
A scaled Gaussian distribution\textcolor{black}{---also called compound Gaussian---}is beneficial as it adjusts variance to better fit data, making it more flexible for capturing different scales \cite{delmas2024}. This adaptability helps to reduce the influence of outliers by preventing them from disproportionately affecting the model. Namely, a vector $\xv$  following  a scaled Gaussian distribution admits the   stochastic representation:
$\xv \isdef \sqrt{\tau}^{-1} \nv$
in which $\isdef$ stands for equal in distribution, the so-called texture parameter $\tau$ is a strictly positive random variable, $\tau\sim f(.;\psiv)$ \textcolor{black}{with unknown $\psiv$, and $\nv\sim \mathcal{N}(\muv,\Sigmab)$ stands for the speckle\footnote{\textcolor{black}{In this setting, $\thev$ includes $\psiv$. The case of an assumed known $\psiv$ can lead to a mismatch scenario, which is illustrated in \cite{mennad2018slepian}.}}.}

In the presence of missing values, a commonly used trick is to apply the chain rule formula in order to reduce it to a classic case, which is the Gaussian distribution presented above. In this case the complete data reads
$\{{\xbo}_{,j},{\xbm}_{,j}, \tau_j\}_j$ and the surrogate function reads
$$
  Q(\thev;\thev^{(k)}) = 
\sum_j  \mathbb{E}_{\tau_j\sim f(\tau_j|{\xbo}_{,j};\thev^{(k)})}[
   \mathbb{E}_{{\xbm}_{,j}\sim f({\xbm}_{,j}|\tau_j,{\xbo}_{,j};\thev^{(k)})}
   [
\log f({\xbm}_{,j},{\xbo}_{,j}|\tau_j;\thev)   ]]
$$
in which the inner expectation $\mathbb{E}_{{\xbm}_{,j}\sim f({\xbm}_{,j}|\tau_j,{\xbo}_{,j};\thev^{(k)})}[
\log f({\xbm}_{,j},{\xbo}_{,j}|\tau_j;\thev)]$ can be derived using the same methodology as in the Gaussian case. 
The outer expectation reduces to the computation of \linebreak
$\mathbb{E}_{\tau_j\sim f(\tau_j|{\xbo}_{,j};\thev^{(k)})}[\tau_j ]$ and $  \mathbb{E}_{\tau_j\sim f(\tau_j|{\xbo}_{,j};\thev^{(k)})}[\tau_j^{-1}]$, which is generally known (or at least easily approximated \cite{cherifi2025}). For example, if $\tau$ is distributed according to an inverse-gamma distribution, it leads to a Student's $t$-distribution for $\xv$.

\subsubsection*{The elliptical symmetric distribution case}
Recently, elliptically symmetric distributions have become popular in signal processing \cite{delmas2024} because they extend the multivariate normal distribution while preserving useful properties such as  ellipsoidal contours. They provide more flexibility, especially for data with outliers or heavy tails, as seen in fields like finance, radar and array processing, biomedical applications, computational imaging, etc.
These distributions also remain mathematically tractable, allowing an easy adaptation of many statistical methods originally designed for the normal case. Furthermore, this makes them valuable for robust modeling and analysis when normality assumptions are too limiting. Consider an elliptical symmetric distributed random variable $\xv\sim ES(\muv,\Sigmab,g)$ in which   $g$ represents the so-called density generator. The associated probability density function reads
    $$f(\xv;\thev) =  \det(\Sigmab)^{-\frac{1}{2}}  g( (\xv-\muv)^\T \Sigmab^{-1} (\xv-\muv)).$$
    One can notice that for different functions $g(\cdot)$, under certain mild conditions, we obtain a specific distribution (see Table.~\ref{Table:CES} for some examples). With missing values, the E-step of the EM algorithm reduces to
\begin{equation}
\label{Q_ES}
     Q(\thev;\thev^{(k)}) \propto -\frac{n}{2} \log \det(\Sigmab)
  + \sum_j   \mathbb{E}_{{\xbm}_{,j}\sim f({\xbm}_{,j}|{\xbo}_{,j};\thev^{(k)})}[ \log g( (\xv_j-\muv)^\T \Sigmab^{-1} (\xv_j-\muv)) ].
\end{equation}

\begin{table}[h!]
\centering
\begin{tabular}{ c | c | c | c | c }
& Gaussian & Student's $t$ &  Generalized Gaussian &    K-distribution  \\
&   & {\scriptsize  $\nu$ degree of freedom}  &  {\scriptsize $s,b$  shape and scale} & {\scriptsize $\nu$ shape}     \\ \hline
 $g(r)$ & $\frac{\exp(-\frac{r}{2})}{(2\pi)^{n/2}}$ &  $\frac{ (1+r/\nu)^{-\frac{n+\nu}{2}} \Gamma(\frac{\nu+n}{2})}{(\nu\pi)^{\frac{n}{2}} \Gamma(\nu/2)}$ &  $\frac{s \exp(-\frac{r^s}{2^sb}) \Gamma(n/2)}{(2\pi)^{n/2} b^{n/2s} \Gamma(n/2s)}$
 & $\frac{\nu^{n/2}(2\nu r)^{(2\nu -n)/4}}{2^{\nu-1}\pi^{n/2}\Gamma(\nu)} K_{\nu-n/2}(\sqrt{2 \nu r})$ \\ \hline
 \end{tabular}
   \caption{Examples of common \textcolor{black}{elliptical symmetric} distributions ($\Gamma(\cdot), K_{.}(\cdot) $ denote the Gamma function and the  modified Bessel function of the second kind, respectively). \label{Table:CES}}
 \end{table}

\subsubsection*{EM's  stochastic variants}
The expectation in (\ref{Q_ES}) is not always feasible, especially for a general form of the density generator. If the E-step is intractable (e.g., due to a complex density generator function), a natural choice is to use a stochastic variant of the EM algorithm. Namely, instead of computing the full expectation, it is approximated using random samples, often via Monte Carlo methods. This enables efficient parameter estimation while maintaining convergence properties under certain conditions. Several stochastic versions of the EM algorithm have been developed to address such kind of situation (see Table~\ref{Stochastic_EM_variants}). Stochastic EM  (SEM) replaces the full E-step with a sample-based approximation, using a single or a few random draws from the conditional distribution of the \textcolor{black}{missing observations}. This allows the algorithm to handle large or complex datasets when computing the exact expectation is impractical.
Monte Carlo EM (MCEM) improves this by using Monte Carlo sampling to approximate the E-step with multiple samples, providing a \textcolor{black}{more accurate estimate of the expectation (one can consider Metropolis-Hasting \cite{cherifi2025}  or Gibbs sampling \cite{liu2019parameter}). The MCEM is especially useful when analytical solutions are unavailable but sampling is feasible \cite{mclachlan2008algorithm}.
Finally, stochastic approximation EM (SAEM) goes further by combining stochastic updates with a smoothing mechanism, updating the expectation incrementally over iterations as $$Q(\thev;\thev^{(k+1)}) = Q(\thev;\thev^{(k)}) +\gamma_k \left[ \log f(\Xbo,\Xbm^{(k)};\thev) - Q(\thev;\thev^{(k)})  \right]$$ s.t. $\Xbm^{(k)} \sim f(\Xbm|\Xbo;\thev^{(k)})$ and $0<\gamma_k<1$ . This leads to better stability and convergence, even when using fewer samples at each iteration \cite{little2019,delmas2024}.} 

\begin{table}[h!]
\centering
\begin{tabular}{l|p{6.5cm}|p{7.5cm}}
 & {E-Step Description} & {Analytical Equation}   \\
\hline
{EM} & Computes the exact conditional expectation of the complete-data log-likelihood. &
%
$ Q(\thev;\thev^{(k)}) = \mathbb{E}_{\Xbm\sim f(\Xbm\vert\Xbo;\thev^{(k)})}[ \log f({\Xbo},{\Xbm};\thev) ]$ 
\\
\hline
{SEM} & Approximates the E-step using a single   sample  from the posterior. &
%
$ Q(\thev;\thev^{(k)}) \approx  \log f(\Xbo,\Xbm^{(k)};\thev) $\hfill\hfill\linebreak where
$\Xbm^{(k)} \sim f(\Xbm|\Xbo;\thev^{(k)})$ 
\\
\hline
{MCEM} & Uses multiple samples to estimate the expectation in the E-step. &
$ Q(\thev;\thev^{(k)}) \approx  \frac{1}{K} \sum_{c=1}^K \log f(\Xbo,\Xbm^{(k),c};\thev) $ \hfill\hfill\linebreak where
$\Xbm^{(k),c} \sim f(\Xbm|\Xbo;\thev^{(k)})$
\\
\hline
{SAEM}   & Uses a recursive update combining current and past estimates. &
%
$Q(\thev;\thev^{(k+1)}) = Q(\thev;\thev^{(k)}) +\gamma_k \left[ \log f(\Xbo,\Xbm^{(k)};\thev) - Q(\thev;\thev^{(k)})  \right]$ \hfill\hfill\linebreak$\Xbm^{(k)} \sim f(\Xbm|\Xbo;\thev^{(k)})$, \qquad $0<\gamma_k<1$ 
\\
\hline
\end{tabular}
\caption{Comparison of E-step formulations in EM variants \cite{mclachlan2008algorithm}}
\label{Stochastic_EM_variants}
\end{table}

\textcolor{black}{
The analysis above remains valid when the mechanism is said to be ignorable (c.f., definition \ref{def:ignorable}). Nevertheless, inference with ignorable mechanisms can be limited  in many real-world signal processing  applications. If the mechanism is  non-ignorable, ignoring the relationship between the missingness and the unobserved data  inevitably leads to biased parameter estimates and fundamentally invalid conclusions. In the following we consider the case of non-ignorable mechanism in the estimation process. }

\subsection*{Statistical inference with non-ignorable mechanism}

As discussed \textcolor{black}{in the section \textit{Basic concepts of missing data}}, in the case of MNAR \textcolor{black}{data}, considering only the observed likelihood can dramatically bias the estimation. 
In this situation, it is necessary to consider the joint distribution $f(\Xbo,\Mb;\thev,\phiv)$. Two important distinctions arise, depending on whether the missingness mechanism is non-ignorable but known (as in censored data with a known threshold, e.g., due to a sensor’s operating range) or non-ignorable and unknown, in which case $\phiv$ must be estimated jointly with $\thev$.
In the following, we focus on the latter scenario, under the assumption of independent observations \textcolor{black}{(to simplify calculations and reading)}, i.e., $f(\Xb,\Mb;\thev,\phiv)=\prod_i \prod_j f(X_{i,j},M_{i,j};\thev,\phiv)$.  
Several models can be used to infer under non-ignorable mechanism depending on the choice of target parameters. In the following, we consider the so-called \textit{selection} model since it is more appropriate in the SP context, the marginal distribution of the data being of interest. Specifically, the selection model reads 
\begin{equation}\label{eq:MNARselectionmodels}
  f(X_{i,j},M_{i,j};\thev,\phiv)= f(X_{i,j};\thev) f(M_{i,j}| X_{i,j};\phiv).
  \end{equation}
  This requires external knowledge of the data model (e.g., a regression model for observed and missing outcomes) and the missingness mechanism (e.g., a logistic or probit model for the missingness probabilities). A natural choice of the conditional distribution of the mask  is
$$f(M_{i,j}|X_{i,j};\phiv)= (h(X_{i,j};\phiv))^{M_{i,j}}  (1-h(X_{i,j};\phiv))^{1-M_{i,j}}$$
where $h(\cdot)$ is any specific function.  

In the context of MNAR (specifically, considering the 
joint distribution $f(\Xbo,\Mb;\thev,\phiv)$), the MLE is impractical and obtaining  closed-form expression of $Q(\thev,\phiv;\thev^{(k)},\phiv^{(k)})$ in the EM algorithm is rarely possible \cite{mclachlan2008algorithm}.  
Again, stochastic variants of the EM (see Table.~\ref{Stochastic_EM_variants}) are therefore commonly employed in this scenario. As an example, the SEM reads
\begin{itemize}
\item \underline{SE-step:} 
\begin{align*}
    Q^{\rm SEM}(\thev,\phiv;\thev^{(k)},\phiv^{(k)})  
= \sum_i\sum_j & 
M_{i,j}(\log f(X_{i,j};\thev) + \log f(M_{i,j}|X_{i,j};\phiv) )
 \\ &   + 
(1-M_{i,j})
(\log f(X_{i,j}^{(k)};\thev) + \log f(M_{i,j}|X_{i,j}^{(k)};\phiv) )
\end{align*}
with $X_{i,j}^{(k)}\sim f(X_{i,j}|M_{i,j};\thev^{(k)},\phiv^{(k)}).$
    \item \underline{M-step:} On the other hand, the M-step for the SEM
     reduces to
    \begin{align*}
        \thev^{(k+1)} & =\argmaxx_{\thev \in \Omega_{\thev}}  \sum_i\sum_j 
M_{i,j}\log f(X_{i,j};\thev)  
 + (1-M_{i,j})
\log f(X_{i,j}^{(k)};\thev),   \\
         \phiv^{(k+1)} & =\argmaxx_{\phiv \in \Omega_{\phiv}} \sum_i\sum_j 
M_{i,j} \log f(M_{i,j}|X_{i,j};\phiv) 
 + (1-M_{i,j})
 \log f(M_{i,j}|X_{i,j}^{(k)};\phiv).
    \end{align*} 
\end{itemize}

\textcolor{black}{
We demonstrated that stochastic EM variants can handle non-ignorable missingness, delivering valid parameter estimates where standard EM fails. Their main drawback is the computational cost of sampling from the missing data’s posterior distribution. However, recent advances mitigate this by approximating the distribution in the E-step, cutting computational overhead without sacrificing statistical accuracy \cite{allassonniere2021new}.}

\subsection*{Estimation with prior information}

\textcolor{black}{ In this section, we focus  on  parameter estimation in the presence of incomplete observations, leveraging two distinct types of \textit{prior information}.
Specifically, the proposed framework seeks the estimation of  the parameter ${\boldsymbol{\theta}}$ by minimizing a cost function that balances data fidelity and parameter constraints, which reads
\begin{equation}
    \label{prior_information}
    \hat{\boldsymbol{\theta}}= \argminn_{\boldsymbol{\theta} \in \Omega_{\thev}} d(\boldsymbol{\theta},\Xbo) + \lambda R(\boldsymbol{\theta})
\end{equation}
where $\Omega_{\thev}$ is the parameter space. Problem~\eqref{prior_information} encompasses two critical estimation paradigms widely used in the SP community. The first uses \underline{statistical priors}: by defining the distance $d(\boldsymbol{\theta},\Xbo)$ as the negative log-likelihood and $R(\boldsymbol{\theta})$ as the negative log-prior, the framework recovers the EM-MAP estimator. The second approach leverages \underline{structural priors}, constraining or parameterizing unknown parameters and/or using a structural regularizer ${R}(\thev)$ (e.g., nuclear norm). This framework extends to probabilistic PCA (PPCA) and subspace tracking, two crucial approaches for recovering (dynamic) low-dimensional signals.} 


\textcolor{black}{
\subsubsection*{1) Statistical prior}
In SP applications, the statistical prior is generally modeled by a prior on the unknown parameter $\thev$, denoted by 
$f(\thev)$. As an example, in \textbf{radar and surveillance} applications, the a priori $f(\thev)$ can be centered on the expected angle, while in \textbf{array processing} applications, the Bernoulli-Gaussian mixture distribution can be an effective choice for $f\thev)$ when the number of active sources is small and unknown (leading to a sparse scenario \cite{barbault2025lemur}).
Taking into account the statistical prior with missing value can be efficiently solved  using the EM-MAP (maximum a posteriori). 
Namely, the MAP reads}
 $$\hat{\thev}_{\rm MAP}
    = \argmaxx_{\thev \in \Omega_{\thev}} \sum_j \log \int  f({\xbo}_{,j},{\xbm}_{,j}\vert\thev)   d{\xbm}_{,j}  +  \log   f(\thev),$$
    in which $ f(\thev) $ denotes the prior information. 
    \textcolor{black}{ In this case, we can notice that the E-step of the EM algorithm remains the same (the expectation is with respect to the latent variable only), whereas the M-step reads as (\ref{prior_information}), where
    $ d(\boldsymbol{\theta},\Xbo)$ reduces to the surrogate $Q(\cdot)$ and  the regularization $R(\boldsymbol{\theta})$ represents the a priori $f(\thev)$ with
    $\lambda=1$. This leads to}
\begin{equation}
\label{EM_MAP}
       {\thev}_{\rm MAP}^{(k+1)} =\argmaxx_{\thev \in \Omega_{\thev}} Q(\thev;\thev^{(k)}) +
   \log   f(\thev).
\end{equation}
Due to the presence of the a priori,   the M-step might  be difficult to compute. 
 In classical EM, the M-step involves a full joint maximization over all parameters, which can become computationally expensive or even intractable for complex models. The ECM (Expectation Conditional Maximization) algorithm addresses this by decomposing the M-step into a series of simpler conditional maximizations over parameter blocks. The ECME (ECM Either) extends this idea by replacing some of these conditional steps with direct maximization of the observed-data log-likelihood, often resulting in faster convergence. Lastly, the GEM (Generalized EM) algorithm relaxes the requirement of full maximization entirely, requiring only that the surrogate function increases at each iteration
 \cite{mclachlan2008algorithm}. This flexibility allows for the use of partial updates or gradient-based methods when exact optimization is not feasible. These variants are particularly valuable in high-dimensional, constrained, or non-convex problems where the classical EM algorithm may not be practical \cite{little2019}.

\textcolor{black}{
\subsubsection*{2) Structural prior}
Many \textcolor{black}{estimation algorithms} can benefit from an available \textit{structural prior knowledge} pertaining to the unknown statistical parameter.} This knowledge can be obtained{\color{black}, for example,} from physical considerations on the observed phenomenon, or by taking advantage of dependencies between variables, which is often the case with large-scale, high-dimensional data.
  \textcolor{black}{In the missing data case, 
$d(\boldsymbol{\theta},\Xbo)$ can be replaced  by the surrogate function $Q(\cdot)$ or by a parametric model fitting. This translates to the following constrained optimization formulation: }

{\color{black}
\begin{align} \label{eq:m_step_structured} 
   \begin{array}{rcl}
      \hat{\boldsymbol{\theta}} =\!\!\!\!    &\underset{\thev \in \Omega_{\thev}}{\mathop{ \mathsf{argmin}}}  & d(\boldsymbol{\theta},\Xbo) + \lambda R(\boldsymbol{\theta}) \\
   & \subto & \Omega_{\thev} = \left\{ \thev = \mathcal{F}(\boldsymbol{\phi}), \boldsymbol{\phi} \in \mathbb{R}^q  \right\},
   \end{array}
\end{align}
where $\mathcal{F(\cdot)}$ is a known one-to-one mapping from $\boldsymbol{\phi}$ to $\thev$ \footnote{In an alternative formulation, the unknown parameter $\thev$ can be reparameterized such that $\hat{\boldsymbol{\thev}} =  \mathcal{F}( \hat{\boldsymbol{\phi}})$, subject to $\hat{\boldsymbol{\phi}} = \underset{\boldsymbol{\phi} \in \mathbb{R}^q}{\mathop{ \mathsf{argmax}}} \ \ d( \mathcal{F}(\boldsymbol{\phi}),\Xbo) + \lambda R( \mathcal{F}(\boldsymbol{\phi})).$}.} \textcolor{black}{To illustrate this, we consider two popular applications of this framework in signal processing : \textbf{PPCA}  and \textbf{subspace tracking}.} 
 
\textcolor{black}{
Let us first consider the PPCA formulation. In this setting, the covariance matrix is assumed to be the sum of a low-rank component plus an identity matrix, which can be modeled by the well-known factor model \cite{ruppert2011}.} 
{\color{black} Here, $\thev$ reduces to the covariance matrix $\Sigmab$ and $\boldsymbol{\phi}=\left\{ \sigma^2, \Rb\right\}$ such that
\begin{equation} \label{eq:lowrank_structure}
\Omega_{\thev} = \Big\{   \Sigmab =
\mathcal{F} (\boldsymbol{\phi}) \stackrel{\text{def}}{=} \sigma^2\Ib_p + \Rb, 
  \ \ \Rb \succeq \bm{0}, \ \  \rank(\Rb) \le r, \ \ \sigma^2 > \sigma_{\textsf{known}}^2 \Big\},
\end{equation}
where $\Ib_p$ denotes the $p$-dimensional identity matrix, $\sigma$ is a strictly positive scalar, {\color{black} $\sigma_{\textsf{known}}^2\ge0$ is a lower bound on $\sigma^2$, and $\Rb$ is factored as $\Ub\Sb\Ub^\T$, with $\Ub$ being a $p\times r$ matrix and $\Sb$ being a $r\times r$ diagonal matrix}.
The solution to \eqref{eq:m_step_structured} with structure~\eqref{eq:lowrank_structure} can be obtained via the EM algorithm: at step $k+1$, it is given by }
\begin{align*}\hat{\Sigmab}^{(k+1)} = \hat{\sigma}^{2(k)}\Ib_p + \Ub^{(k)}\hat{\Lamb}^{(k)}\Ub^{\T(k)} 
\quad
\textrm{with} \quad
\begin{cases} 
\hat{\sigma}^{2(k)} = \frac{1}{p-r}\sum_{i=r+1}^p\lambda_i^{(k)} \\
\hat{\Lamb}^{(k)} = \diag(\lambda_1^{(k)} - \hat{\sigma}^{2(k)}, \dots, \lambda_r^{(k)}-\hat{\sigma}^{2(k)}),
\end{cases}
\end{align*}
where $\hat{\bm{\Sigma}}^{(k)} \overset{\text{EVD}}{=} 
\sum_{i=1}^{p}\lambda_i^{(k)}\uv_i^{(k)}\uv_i^{\T(k)}$,  $\; \lambda_i^{(k)}<\dots<\lambda_p^{(k)}$ are the eigenvalues of $\hat{\bm{\Sigma}}^{(k)}$ and $\uv_1^{(k)}, \dots, \uv_p^{(k)}$ are the corresponding eigenvectors contained in $\Ub^{(k)}$. \textcolor{black}{Note that $\Sigmab^{(0)}$ can be initialized using the sample covariance matrix (SCM) computed from fully observed samples $\xv_i$ in the data matrix $\Xb$.}

\noindent 
\textcolor{black}{
Structure~\eqref{eq:lowrank_structure}, along with its constrained optimization problem~\eqref{eq:m_step_structured}, is relevant for several SP applications. For example, in \textbf{radar processing}, the observations are collected in $\Xb$, which is often incomplete due to transmission-reception failures or sensor failures. In this case, $\Sb$ represents the   diagonal source covariance matrix to be estimated, $\Ub$ the array manifold matrix and $\sigma^2$   the thermal  power of the   noise. \eqref{eq:lowrank_structure} can also be used to perform {adaptive beamforming} with monotone missing values in $\Xb$, where the covariance matrix must be estimated from incoming data to construct an adaptive beamformer (c.f., \cite{aubry2021} in which the authors   explored other useful covariance structures for radar processing applications). {\color{black} In \textbf{image analysis}, structure~\eqref{eq:lowrank_structure} has been used in \cite{hippert2022} to perform covariance-based clustering of hyperspectral data with a scaled-Gaussian distribution. In this setting, the data represents $n$ observed pixels over $p$ bands. It can be incomplete due to dead and/or saturated pixels, and to band dropout, which leads to a general pattern of missing values.}}

\textcolor{black}{Let us now consider the extension of PPCA in the \textit{streaming} scenario, namely, robust subspace tracking (RST), which adaptively estimates and updates the subspace of a signal as it evolves over time \cite{thanh2021a}. It concerns a wide range of applications involving high-dimensional data and time-varying systems, such as adaptive filtering, direction-of-arrival estimation, or blind source separation\textcolor{black}{, among others}. In the past fifteen years, many techniques have flourished to adapt RST to missing data.
Specifically, in scenarios where the data exhibit a certain level of redundancy, such as when complete high-dimensional data can be well approximated by their projection onto a lower-dimensional subspace, incomplete data become particularly valuable, as they may still contain sufficient information thanks to this redundancy \cite{Balzano2018,thanh2021b}. } \textcolor{black}{In this context,  the true time variate (unknown) signal  $\xv_t \in \mathbb{R}^p$ is assumed to belong to an {\color{black}  \textit{unknown}} low dimensional subspace $\Ub \in \mathbb{R}^{p\times r}$ ($r\ll p$), i.e.,  $\xv_t = \Ub \wv_t$, where $\wv_t$ is an \textit{unknown} weight vector. }
\textcolor{black}{ Many of the subspace tracking algorithms seek to optimize a recursively time variant updated loss function 
$F_t(\cdot)$, e.g., the squared projection loss onto the subspace, and can be cast into (\ref{eq:m_step_structured}), as}
\begin{equation}
\label{eq:FU}
    F_t(\Ub) = \alpha_t F_{t-1}(\Ub) + \beta_t \big\norm
    \mv_t\odot ({\yv}_{t} - \Ub\wv_t)\big\norm_2^2,
\end{equation}
where ${\yv}_{t}$ is the observed vector, $\mv_t$ the associated mask, $\alpha_t$ and $\beta_t$ control the trade-off between convergence rate and tracking adaptability to subspace changes, and $\Ub$ is typically updated using the previous estimate. The minimization framework associated to this problem for the estimation of $\Ub$ is based on the following general objective, \textcolor{black}{which is a specific case of   (\ref{prior_information})}:
\begin{equation}
\label{eq:ST_cost} 
\Ub_t = \underset{\boldsymbol{\Ub}}{\argminn} \;     F_t(\Ub) + \rho g(\Ub)
\end{equation}
{\color{black} where $g(\Ub)$ is an optional regularization function on $\Ub$ (e.g., a sparsity promoting function in the presence of outliers \cite{thanh2021a}) and $\rho$ controls the regularization level}.
\textcolor{black}{
One way to solve (\ref{eq:FU})-(\ref{eq:ST_cost}), is parallel estimation and tracking by recursive least squares (PETRELS). This approach is a second-order recursive least-squares (RLS) algorithm \cite{chi2013} that employs second-order stochastic gradient descent on the cost function.} At each time $t$, the weight vector $\wv_t$ and subspace $\Ub_t$ {\color{black} are alternatively estimated by minimizing the projection residual on the previous subspace estimate $\Ub_{t-1}$, where $\Ub_0$ is a random subspace initialization. Note that \cite{Balzano2024} have proposed a probabilistic formulation of problem~\eqref{eq:ST_cost}, which requires some assumptions on the data distribution. In this setting, $\Ub_t$ and $\wv_t$ are estimated at time $t$ by maximizing the observed log-likelihood using a stochastic alternating EM approach.

\begin{tcolorbox}[colback=gray!10!white, colframe=gray!50!black, title=Application 3 : Graph Learning for Spatio-Temporal Signals with  Missing Values, sidebyside,sidebyside align=top,lower separated=false, breakable, bicolor,
after={\noindent}]

We earlier discussed graph signals and their numerous applications.
The primary step in graph signal processing is to identify a graphical model that best captures the similarities or dependencies in the signal.
This process is referred to as \emph{graph learning},
for which there are various approaches,
broadly classified into \emph{undirected} and \emph{directed} methods.
An undirected graph    models  bilateral relationships, while a directed graph is used to represent unilateral dependencies. 
Let 
$\Xb$
denote the graph signal matrix  and let $\thev$ be the parameters of the underlying graph model. 
Then the MAP formulation for  estimation of $\thev$ given the data yields:
\begin{align}
\label{eq:Prob_GL}
\begin{split}    
    \hat{\thev}_{\text{MAP}} &= 
    \argmaxx_{\thev\in \Omega_{\thev}}\,\, \log f(\thev \vert\Xb)  =
    \argminn_{\thev\in \Omega_{\thev}}\,\, -\log f(\Xb\vert\thev) - \log f(\thev).  
     \end{split}
\end{align}
This formulation is specified in Table~\ref{table:GL} for two standard graph learning frameworks: the Gaussian Markov random field (GMRF) model \cite{egilmez2017} for undirected graphs and the vector autoregressive (VAR) model\cite{ioannidis2019} for directed graphs. 

The graph learning framework based on \eqref{eq:Prob_GL} assumes complete data statistics, while in practice, observations often contain missing values due to various factors such as sensor failures or measurement errors.
One may consider 
the missing samples as  latent variables and  use EM 
for the estimation of  the graph.
This approach, however, only infers the graph structure, requiring additional processing for  imputation of missing values. 

An alternative approach is to simultaneously estimate the missing values and the graph parameters,
enabling joint graph learning and signal imputation,
while reducing computational complexity. 
Let $\Xb =[\xv_1,\hdots, \xv_n]\in \Real^{p \times n}$ represent the complete (ground-truth) data matrix.
Consider the case where the observed data contains missing values and noise. Replacing the missing values  with $0$, the matrix of the  observations could then be modeled as  $\Yb = \Mb\odot(\Xb + \Nb)$  where the  missing mask $\Mb\in \{1,0\}^{p\times n}$ specifies the missing pattern and  $\Nb\in \Real^{p\times n}$ represents the measurement noise.
Assuming the missing data mechanism is
{\color{black} MCAR,}
the MAP estimation of the graph parameters $\thev$ {\color{black}(see Table \ref{table:GL} for definition)} and the complete data matrix $\Xb$  given $\Yb$ and $\Mb$ is formulated as 
\begin{align}
\label{eq:MAP_graph_missing}
\begin{split}
    \hat{\thev}_{\text{MAP}}, \hat{\Xb}_{\text{MAP}} 
    &= \argmaxx_{\thev\in \Omega_{\thev},\,\Xb } \,\log f(\thev; \Xb\vert \Mb,\Yb), \\
    &= \argminn_{\thev\in \Omega_{\thev},\,\Xb}\, -\log f(\Yb\vert \Xb, \Mb)  - \log f(\Xb\vert\thev) - \log f(\thev).
  \end{split}
\end{align}

Classical  methods for graph learning, however, only capture either spatial correlations through an undirected graph structure (GMRF model), or   temporal dependencies via a directed graph topology (VAR model).  
For spatio-temporal signals (i.e., networked time series data), both spatial and temporal dependencies are crucial. Recent approaches like \cite{javaheri2024a} have therefore introduced a multi-relational graph model comprising both undirected and directed structures to simultaneously represent spatial and temporal dependencies.
This model is described by:
 \begin{align}
     \xv_t =\Ab \xv_{t-1} + \epv_t,\qquad f(\epv_t|\Lb) \propto \sqrt{\mathrm{det}^*\Lb} \exp\left(-\frac{1}{2}\epv_t^\T\Lb\,\epv_t\right).
 \end{align}
Here, $\xv_t$ and $\epv_t$ are  the signal and the innovations at time $t$, respectively; $\Ab$ represents the adjacency matrix of a directed graph capturing temporal dependencies, and  $\Lb$ denotes the Laplacian matrix of an undirected graph encoding spatial correlations.
 The parameters of this model are    $\thev = (\Ab,\Lb)$.

The {spatio-temporal  signal recovery and graph leaning} (STSRGL) method  in \cite{javaheri2024a} (code available at \href{https://github.com/javaheriamirhossein/STSRGL}{https://github.com/javaheriamirhossein/STSRGL})  solves  problem \eqref{eq:MAP_graph_missing} for joint inference of this spatio-temporal model.

In Fig. 
~\ref{fig_GL_directed},  
we compare the STSRGL method with baseline algorithms for
\begin{itemize}
    \item spatial graph learning: {combinatorial graph learning} (CGL) \cite{egilmez2017} method; 
    \item temporal graph inference: {joint inference of signals and graphs} (JISG) \cite{ioannidis2019} method.
\end{itemize}

The evaluation uses  
spatio-temporal data defined over a $10 \times 10$ grid graph 
($p=100$), 
featuring $50\%$ missing values and additive Gaussian noise ($\sigma_n^2=0.01$). As shown in the figures, the STSRGL more accurately identifies both the non-zero support (connections) and the entries (edge weights)  of the graph matrices compared to methods that learn only spatial or temporal graphs.

\centering
   \includegraphics[trim={0 0 0 0},clip,width=0.27\textwidth]{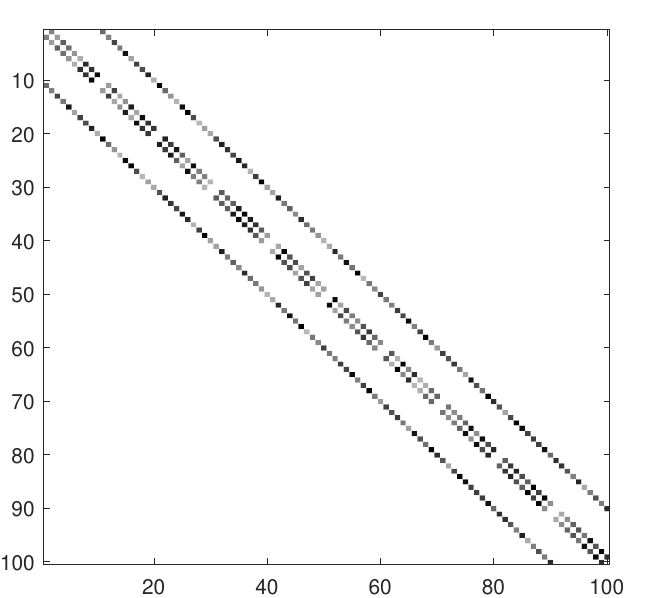}
   \includegraphics[trim={0 0 0 0},clip,width=0.27\textwidth]{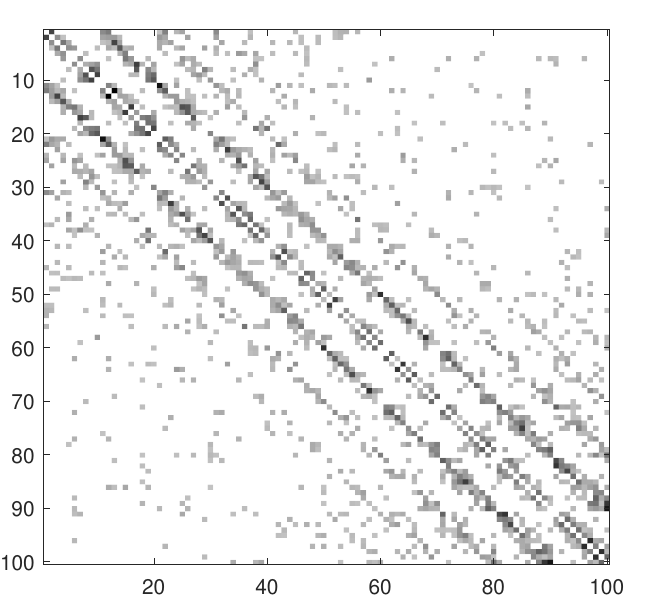}
   \includegraphics[trim={0 0 0 0},clip,width=0.27\textwidth]{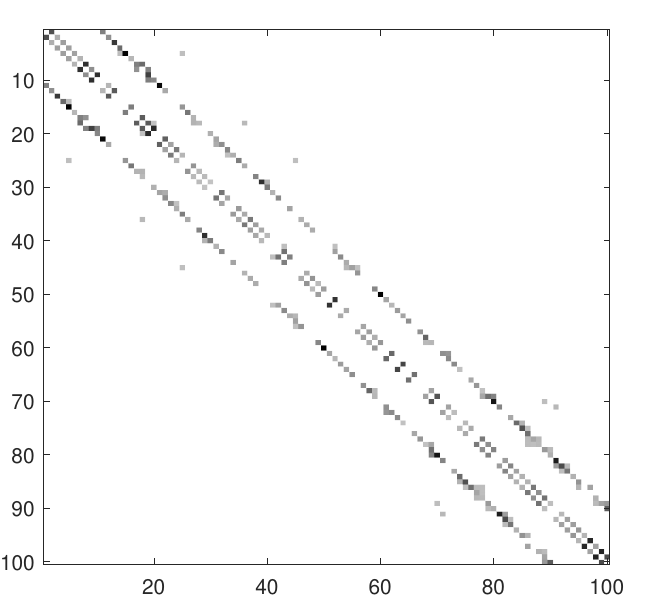}

\includegraphics[trim={0 0 0 0},clip,width=0.27\textwidth]{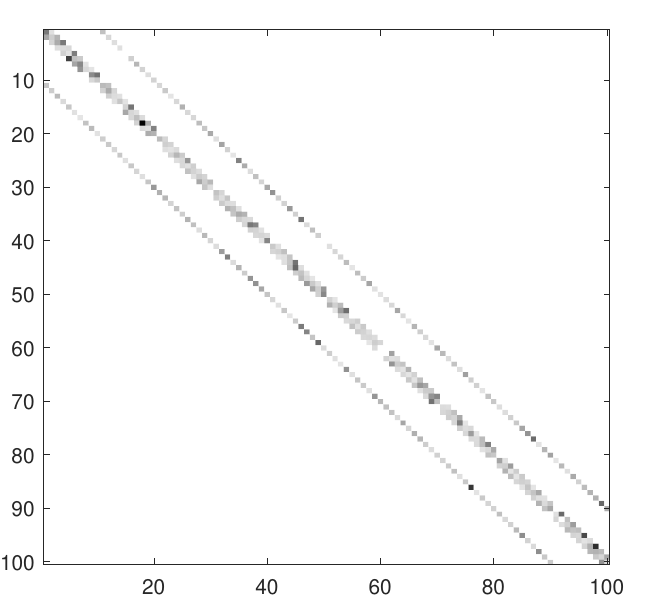}
   \includegraphics[trim={0 0 0 0},clip,width=0.27\textwidth]{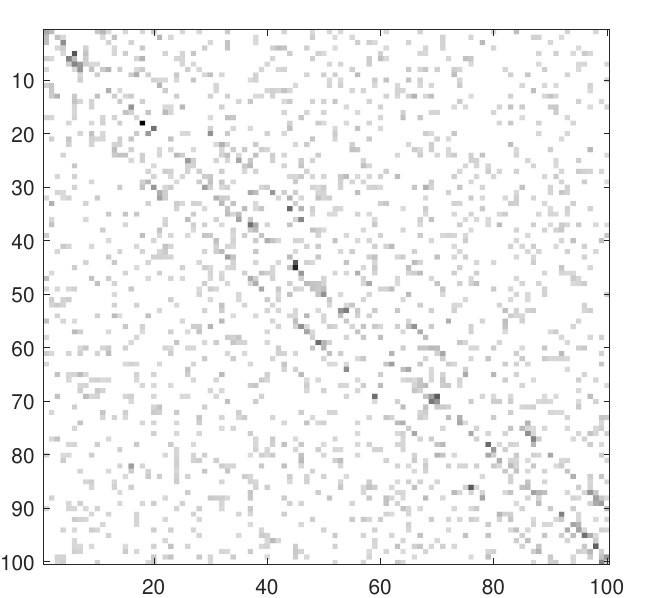}
   \includegraphics[trim={0 0 0 0},clip,width=0.27\textwidth]{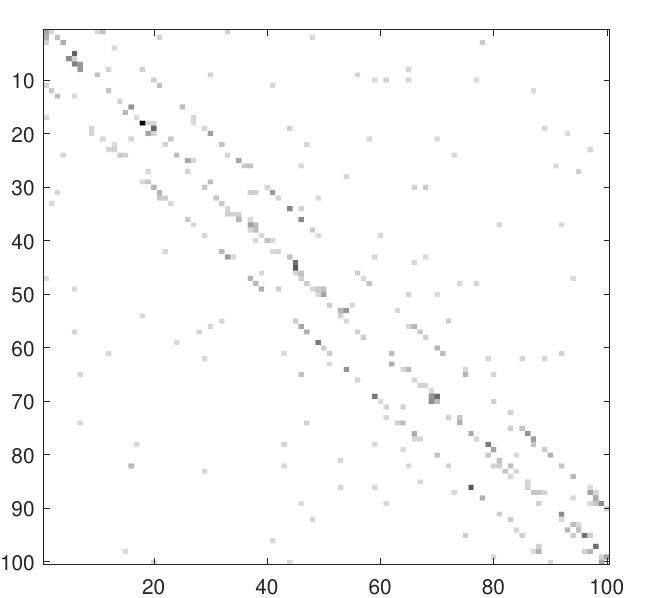}
\captionsetup{font=small} \captionof{figure}{ Visualization of 
$\vert 
L_{i,j}\vert\,\, (i\neq j)$ (top) and $\vert A_{i,j}\vert$ (bottom)
learned  from 
data with  noisy and missing observations. 
Top: the ground-truth (left),  the CGL (middle), and the STSRGL (right)  spatial graphs. 
%
Bottom: the ground-truth (left),  the JISG (middle), and the STSRGL  right)  temporal graphs.  }
\label{fig_GL_directed}

\begin{table}[H]
\hspace*{-10pt}
\centering
	\begin{tabular}{c | c | c | c   } 
    \!\!Graph Type\!\! & Model & \!\!\!\!Parameters $\thev$ \!\!\!\!  & $-\log f(\Xb;\thev)-\log f(\thev)$ \,\,(up to scale) \\
		\hline
		\hline
        \multirow{2}*{\!\!\!\!Undirected\!\!\!\! }   &
         {\color{black}Gaussian} Markov random field   & 
        \multirow{2}* { \hspace*{-5pt}$\Lb$ (Laplacian matrix)} \hspace*{-5pt}& 
        \multirow{2}*{\!\!$\tr\!\left(\!\Lb \displaystyle\sum_{j=1}^n \frac{\xv_j \xv_j^\T}{n}\!\right)\!-\log {\det^\ast} \Lb + \alpha \normop{\Lb}_{1,\text{off}}  $} \\
        & \!\!$f(\xv_j;\Lb) \propto \sqrt{\mathrm{det}^*\Lb} \exp\left(-\frac{\xv_j^\T\Lb\xv_j}{2}\right)\!\!$ & 
        {\color{black}
        \hspace*{-10pt}$\begin{array}{c}
        \Lb\in \PSD, L_{i,{j\neq i}}\leq 0, \\[-5pt]
        \Lb\onev = \zerov
        \end{array}$\hspace*{-10pt}
        }
        & \\
        \hline
        \multirow{2}*{\!\!\!\!Directed\!\!\!\! }   &
         vector auto-regressive    & 
        \multirow{2}*{ \hspace{-15pt} $\Ab$ (adjacency matrix) \hspace{-10pt}} & 
        \multirow{2}*{$\sum_{t=2}^n \normop{\xv_t - \Ab \xv_{t-1}}^2 + \alpha \normop{\Ab}_{1,1}  $} \\
        & $\xv_t = \Ab\xv_{t-1} + \epv_t,\,\, \epv_t\sim \Nc(\zerov, \sigma_\epsilon^2\Ib)$ & $\Ab\in \Real^{p\times p}$ &

    \end{tabular}
     \caption{Two standard  models for unpredicted and directed graph learning.   }
      \label{table:GL}
\end{table}
\end{tcolorbox}

{\color{black}

\paragraph*{Takeaway} In the context of parameter estimation, likelihood-based frameworks, including EM and its stochastic  variants (SEM, MCEM, SAEM), are well-suited for structured problems like covariance matrix estimation, subspace tracking or graph learning. These methods provide rigorous uncertainty quantification and handle missingness mechanism rigorously, but may encounter scalability challenges in high-dimensional regimes. 
Furthermore, utilizing the EM algorithm requires specific expertise in statistical modeling, and the need to tailor each EM variant to the specific application limits the feasibility of having a general, unified software package. To overcome these issues, researchers are increasingly turning to deep learning-based methods for handling missing data, offering a potentially more scalable and less domain-specific approach.
}

\section*{Learning with missing data}\label{sec:learning}


{\color{black}
This section highlights modern learning strategies for handling missing data, a domain still underexplored in SP applications. We encourage SP practitioners to adopt these advanced tools, either by tailoring them to their applications or directly incorporating them into their methodologies.
}

\textcolor{black}{A central question in learning with incomplete data concerns the intended downstream application, which fundamentally shapes the choice of modeling strategy. In this section, we classify these applications into two broad categories: imputation, often coupled with an underlying inference problem, and prediction. These tasks give rise to two major modeling paradigms: \textit{generative} and \textit{discriminative} approaches. We first discuss the imputation task, which typically relies on generative models that aim to learn the full data distribution 
$f(\Xb)$. We then turn to prediction, which is more closely aligned with discriminative methods targeting, {in some cases}, the conditional distribution 
$f(\yv|\Xb)$, where $\yv$ is the target variable\footnote{Note that in this section, $\yv$ does not represent a column of $\Yb=\Mb\odot\Xb$ as introduced in \textit{Basic concepts of missing data}.}, thereby avoiding the need to explicitly model 
$f(\Xb)$. }

\subsection*{Generative approaches for imputation}\label{sec:generative} 


\textcolor{black}{
This section introduces the main families of deep learning-based imputation methods. We first present variational approaches, including VAE (Variational AutoEncoder) and IWAE-based models (Importance-Weighted AutoEncoder), which treat missingness through latent-variable modeling and reconstruct unobserved values by optimizing lower bounds on the incomplete-data likelihood. We then describe adversarial methods, such as GAN (Generative Adversarial Network), which formulate imputation as an adversarial game to produce realistic replacements for missing entries. Finally, we discuss OT–based (Optimal Transport) approaches, which view imputation as learning a mapping from incomplete to complete samples by minimizing a transport cost.}

\textcolor{black}{
Although grounded in different principles (variational inference, adversarial learning, or optimal transport) all these methods can be unified under a common generative perspective: they aim is to construct an imputation operator parameterized by a learnable deep neural network that models the conditional distribution of missing values given the observed data. This reads
\begin{align} \label{eq:deep_unification}
   \hat{\Xb}_{\textsf{m}} & = {I}_{\hat{\boldsymbol{\zeta}}} (\Xbo,\Mb)  \\ \nonumber
   \hat{\boldsymbol{\zeta}} & = \argminn_{\boldsymbol{\zeta}}
   \mathbb{E}
   \{  \mathcal{D}_{\boldsymbol{\zeta}}(\Xbo, \Mb) \},
\end{align}
where $\hat{\Xb}_{\mathsf{m}}$ are the imputed missing values, ${I}_{\hat{\boldsymbol{\zeta}}} (\cdot)$ denotes the imputation algorithm indexed by parameters $\hat{\boldsymbol{\zeta}}$, which are learned using a deep neural network,
and $ \mathcal{D}_{\boldsymbol{\zeta}}(\cdot)$ represents a specific divergence or discrepancy. }

\subsubsection*{Variational autoencoders for imputation}
\textcolor{black}{
Several methods use deep latent variable models to estimate the distribution $f(\xv) = \int
f(\xv|\zv) f(\zv) d\zv$, through a variational bound on the observed log-likelihood, which is typically intractable. Here, $\zv$ is a latent variable following a prior distribution $f(\zv)$. 
The main idea of a variational autoencoder based scheme for missing data  is to posit a conditional variational distribution, through the latent variable $\zv$, that will be a learnable proposal \cite{nazabal2020handling,mattei2019miwae,ipsen2020not,ma2021identifiable}. In this perspective, minimizing the discrepancy  $ \mathcal{D}_{\boldsymbol{\zeta}}(\cdot)$ in (\ref{eq:deep_unification}) reduces to minimizing the Kullback-Leibler (KL) divergence between the true posterior distribution and the variational distribution. 
Specifically, by  denoting the   variational learnable distribution by $\tilde{f}(\cdot)$,  the Missing Variational AutoEncoder bound (MVAE) under MAR and MCAR scenario 
 reads
\begin{align}
    \label{elbo_mvae_missing}
    \mathcal{L}_{\rm MVAE}(\thev,\gamv)& =\sum_{j=1}^n \mathbb{E}_{\zv_{j}\sim \tilde{f}(\zv|{\xbo}_{,j};\boldsymbol{\gamv})}
    \left[ \log \frac{f({\xbo}_{,j}|\zv_{j};\thev)f(\zv_{j})}{\tilde{f}(\zv_{j}|{\xbo}_{,j};\gamv)} \right] \\ \nonumber
    & = \ell(\Xbo; \thev) - {\rm KL}\Big(\prod_{j=1}^n \tilde{f}(\zv_{j}|{\xbo}_{,j};\gamv) || \prod_{j=1}^n  f(\zv_{j}|{\xbo}_{,j};\thev) \Big),
\end{align}
where $\ell(\Xbo;\thev)=\sum_{j=1}^n \log f({\xbo}_{,j};\thev)$ denotes the observed log-likelihood.
$\mathcal{L}_{\rm MVAE}(\thev,\gamv)$ is obtained by considering the generative model as a stochastic mapping, parameterized by $\thev$, from a latent variable $\zv_j$ to each row $\xv_j$ of the data matrix $\Xb$, and a variational distribution
$\tilde{f}(\zv_j|{\xbo}_{,j};\gamv)$ targeting the intractable 
${f}(\zv_j|{\xbo}_{,j};\thev)$, as follows
\begin{equation}\label{eq:MVAE-model}
    \zv_j 
    \sim
    \tilde{f}(\zv_j|{\xbo}_{,j};\gamv)  \ \ \ {\text{and}} \ \ \
    \xv_j  
    \sim
    f(\xv_j|\zv_j;\thev).
\end{equation}
These parameters are learned by a deep neural network known as the \textit{encoder} to estimate $\gamv$ and a deep neural network known as the \textit{decoder} to estimate $\thev$. 
Once $\gamv$ and $\thev$ are learned, several imputation scenarios can be considered as particular case of (\ref{eq:deep_unification}). As an example, using  (\ref{eq:MVAE-model}), the operator 
${I}_{\hat{\thev},\hat{\gamv}} (\Xbo,\Mb)$ can sample first the latent variable 
$\zv_j     \sim   \tilde{f}(\zv_j|{\xbo}_{,j};\hat{\gamv})$ and then sample the imputed missing value as
$\xv_j      \sim     f(\xv_j|\zv_j;\hat{\thev})$. 
}

\subsubsection*{Deep generative modeling by importance-weighted autoencoder}
\textcolor{black}{
The performance of the MVAE have been improved in \cite{mattei2019miwae}.   Namely,  the authors in \cite{mattei2019miwae} proposed a tighter bound by using the data Importance-Weighted AutoEncoder (MIWAE) method to impute MCAR or MAR values. Again, the generative model is described by a stochastic mapping which reads
\begin{equation}\label{eq:MIWAE-model}
    \zv_j 
    \sim
    \tilde{f}(\zv_j|{\xbo}_{,j};\gamv){\,=\,}\mathcal{N}\left(\muv_{\gamv}({\xbo}_{,j}),\Sigmab_{\gamv}({\xbo}_{,j})\right) \ \ \ {\rm and} \ \ \
    \xv_j  
    \sim
    f(\xv_j|\zv_j;\thev)
    =
    \mathcal{N}\left(\muv_{\thev}(\zv_j),\Sigmab_{\thev}(\zv_j)\right).
\end{equation}
In practice, the latter conditional and variational distributions are typically modeled as a Gaussian (for sake of simplicity, we restrict our self to continuous data), with parameters $\gamv$ and
$\thev$ that are learned by a deep neural network (\textit{decoder} and \textit{encoder}) in the same sprit as the MVAE. Nevertheless, unlike the MVAE, the variational lower bound to be maximized reads
\begin{equation}\label{eq:lowerbound_MIWAE}
\mathcal{L}_{\rm MIWAE}^K(\thev,\gamv;{\xbo}_{,j})=\sum_{j=1}^n \mathbb{E}_{\zv_{j}^{(1)},\dots,\zv_{j}^{(K)}\sim \tilde{f}(\zv_{j}|{\xbo}_{,j};\gamv)}\left[\log \frac{1}{K}\sum_{c=1}^K \frac{f({\xbo}_{,j}|\zv_{j}^{(c)};\thev)f(\zv_{j}^{(c)})}{\tilde{f}(\zv_{j}^{(c)}|{\xbo}_{,j};\gamv)}\right],
\end{equation}
with $K\in\mathbb{N}^*$ being the number of draws, and we have $\mathcal{L}_{\rm MIWAE}^K(\thev,\gamv)\leq \ell({\Xbo};\thev)$. Note that $f({\xbo}_{,j}|\zv_{j}^{(c)};\thev)$ is obtained by marginalizing the conditional joint distribution $f(\xv_j|\zv_j;\thev)$ in \eqref{eq:MIWAE-model}. From this, we can notice that the MIWAE with $K=1$ reduces to the MVAE, i.e., 
$ \mathcal{L}_{\rm MIWAE}^1 = \mathcal{L}_{\rm MVAE}
.$
The lower bound is then optimized using stochastic gradient-based algorithms. 
Finally, the imputation step is performed either by sampling from the conditional distribution of the missing variables given the observed ones, or for a single imputation, by computing the following conditional expectation: 
\begin{equation}
\label{eq:miwae_xmgivenxo}
    \mathbb{E}[{\xbm}_{,j}|{\xbo}_{,j}; \thev]
=\int \int {\xbm}_{,j} {\frac{f(\zv_j|{\xbo}_{,j};\thev)}{{\tilde{f}(\zv_j|{\xbo}_{,j};\gamv)}}}{f({\xbm}_{,j}|{\xbo}_{,j},\zv_j;\thev)\tilde{f}(\zv_j|{\xbo}_{,j};\gamv)}d\zv_j \,d{\xbm}_{,j}.
\end{equation}
The encoder $\tilde{f}(\zv_j|{\xbo}_{,j};\gamma)$ being a variational distribution, 
$\mathbb{E}[{\xbm}_{,j}|{\xbo}_{,j}; \thev]$
can be approximated \textcolor{black}{using a specific Monte Carlo technique}, called self-normalized importance sampling \cite[Section 9.2]{mcbook} as 
\begin{equation}
\label{eq:imputation_MIWAE}
    \hat{\bm{x}}_{{\sf m},j} =
    \sum_{c=1}^K w_c {\xv}_{\textsf{m},j}^{(c)}, \ \ \ {\text{in}  \ \text{which} }
\end{equation}
\begin{equation}\label{eq:miwae_sampling}
\text{$w_c=\frac{r_c}{\sum_{c=1}^K r_c}$, $r_c=\frac{f({\xbo}_{,j}|\zv_j^{(c)};\thev)f(\zv_j^{(c)})}{{\tilde{f}(\zv_j^{(c)}|{\xbo}_{,j};\gamv)}}$, $({\xv}_{\textsf{m},j}^{(c)},\zv_j^{(c)}) \overset{\mathrm{iid}}{\sim} {f({\xbm}_{,j}|{\xbo}_{,j},\zv_j;\thev)\tilde{f}(\zv_j|{\xbo}_{,j};\gamv)}$}.
\end{equation}
}


\textcolor{black}{
The limitation of the MIWAE is that it is only adapted for the MAR and MCAR scenario and fails when the mechanism is not ignorable. A recent extension, called not-MIWAE, adapts the framework MIWAE to the MNAR setting \cite{ipsen2020not}. Since the missing-data mechanism is non-ignorable, we cannot rely solely on the observed-data likelihood $f({\xbo}_{,j};\thev)$. 
Concretely, the main change occurs in the variational lower bound \eqref{eq:lowerbound_MIWAE}: the joint distribution (the numerator of the terms in the sum) now involves the missing-data mechanism $f({\xbo}_{,j}, \mv_j, \zv_j ; \thev, \phiv)=f(\mv_{j}|{\xbo}_{,j},{\xbm}_{,j};\phiv)f({\xbo}_{,j}|\zv_{j};\thev)f(\zv_{j}).$ While only the latent variable $\zv$ need to be drawn in \eqref{eq:lowerbound_MIWAE}, both the latent variable $\zv$ and the missing part ${\xbm}$ will be drawn for computing the lower bound of not-MIWAE, with the same distribution as in \eqref{eq:miwae_sampling}.}

\subsubsection*{Generative adversarial network for imputation}
\textcolor{black}{
While MIWAE tackles imputation by explicitly approximating the data distribution via the Evidence Lower Bound, GANs offer an alternative paradigm based on implicit modeling \cite{goodfellow2014generative}. Let us describe GAIN, an architecture dedicated to the imputation task \cite{yoon2018gain}.} 

\textcolor{black}{
Without missing data, a Generator ($G$) tries to create realistic data from random noise to fool a Discriminator ($D$), while $D$ tries to distinguish between real data and fake data. Both $G$ and $D$ are fully
connected neural nets. In the missing data setting, 
we only have a incomplete dataset. Therefore, GAIN is defined such that
$G$ observes the real components of the data and attempts to impute the missing components. On the other hand, $D$ attempts to determine which components of a completed vector were actually observed and which were imputed by  $G$.
Specifically,  $G$ takes the observed part of the data $\Xbo$ and the mask $\Mb$ as input. Its outputs reads
\begin{equation}\label{eq:GAIN}
\hat{\xv}_j =  (\mathbf{1}_{n} - \mv_j) \: \odot G({{\xbo}_{,j}}, \mv_j) + \: \mv_j \:  \odot \xv_j
\end{equation}
where $\mathbf{1}_{n}$ is a vector of $n$ components equal to one.
The final imputed vector, $\hat{\xv}_j$, is constructed by keeping the originally observed values and replacing the missing ones with the Generator's output. The Discriminator $D$ receives the completed vector $\hat{\xv}_j$. The main difference between the original GAN and GAIN is that the output is a single scalar (indicating a status real or fake), whereas the GAIN Discriminator outputs a probability vector of the same dimension as the data (i.e., $[D(\hat{\xv}_j)]_i$, represents the probability that the $i$-th element was {observed}  rather than imputed).
In addition to the Generator and Discriminator, the authors in \cite{yoon2018gain} added a hint variable, $\boldsymbol{H}$. 
Specifically, the core function of the hint random variable is to modulate the amount of knowledge the Discriminator possesses about which data points are missing (i.e., the knowledge of $\Mb$). The authors theoretically establish that this control is critical. Without providing $ {D}$ with an adequate hint (or simply if $\boldsymbol{H}$ were omitted entirely) the adversarial game would permit the Generator to learn several different distributions that would all be optimal with
respect to $D$.}

The training is governed by a minimax game involving two loss functions, which are straightforwardly cast into the general formula (\ref{eq:deep_unification}). Specifically, the generator and discriminator are trained simultaneously in an adversarial process with a minimax optimization problem, as follows:
$$\maxx_{G} \minn_{D} \: -\sum_{i=1}^p \sum_{j=1}^n M_{i,j} \log\hat{M}_{i,j} + (1-\hat{M}_{i,j})\log (1-\hat{M}_{i,j}),$$
with $\hat{M}_{i,j}=D({\hat{X}}_{i,j},H_{i,j})$. 
The discriminator aims to accurately recover the missing-data mask (by minimizing the cross-entropy loss), while the generator seeks to maximize this loss by producing imputations that are indistinguishable from observed values, thereby deceiving the discriminator.
Once training is complete, the generator $G$ is used to impute missing values by applying it to incomplete data corrupted with random noise. 

\subsubsection*{Optimal transport}
\textcolor{black}{
While VAEs focus on maximizing conditional likelihood and GANs on learning via discrimination, methods based on OT adopt a fundamentally different approach. They aim to find the most efficient transformation to align the distribution of the imputed data with the distribution of the 
{observed} data. OT is a mathematical framework used to calculate the most efficient way to transform one distribution of mass into another and the Wasserstein distance represents the minimum amount of effort required for this move. If one wants to move from distribution $\mu$  to distribution $\nu$, the squared Wasserstein distance reads \cite{muzellec2020missing}
\begin{equation}\label{eq:opti}
W_2^2(\mu, \nu) = \inf_{\pi \in \Pi(\mu, \nu)} \int  \| \hat{\boldsymbol{X}}_L   - \hat{\boldsymbol{X}}_K \|^2_F d\pi(\hat{\boldsymbol{X}}_L, \hat{\boldsymbol{X}}_K)
\end{equation}
in which $\Pi(\mu, \nu)$ represents the set of all joint probability measures (transport plans) with marginals $\mu$ and $\nu$, and $\hat{\boldsymbol{X}}_L$ and $\hat{\boldsymbol{X}}_K$ represent two batches extracted from the imputed 
matrix $\hat{\boldsymbol{X}}$. The set of indexes $L$ and $K$ are sampled randomly from $\{1,\dots,n\}$, such that $L \cap K = \varnothing$ and $|L| = |K| < n$ (for example, if $L=\{1,2,3\}$, $\hat{\Xb}_L$ contains the three first samples (columns) of $\hat{\Xb}$, that is $\hat{\Xb}_L=[\xv_1,\xv_2,\xv_3]$). 
The Wasserstein distance is nevertheless computationally expensive and is not strictly differentiable everywhere, which poses a problem for gradient-based optimization. To solve this, \cite{muzellec2020missing} utilized the following OT cost:
$$OT_{\varepsilon}(\mu, \nu) = \inf_{\pi \in \Pi(\mu, \nu)} \int  \| \hat{\boldsymbol{X}}_L   - \hat{\boldsymbol{X}}_K \|^2_F d\pi(\hat{\boldsymbol{X}}_L, \hat{\boldsymbol{X}}_K) - \varepsilon H(\pi),$$
where $H(\pi)$ is the entropy of the transport plan. 
This formulation makes the distance differentiable but is biased. To correct this, one can employ the Sinkhorn Divergence, denoted as $\mathcal{S}_{\varepsilon}$:
$$\mathcal{S}_{\varepsilon}(\mu, \nu) = OT_{\varepsilon}(\mu, \nu) - \frac{1}{2}OT_{\varepsilon}(\mu, \mu) - \frac{1}{2}OT_{\varepsilon}(\nu, \nu).$$
By minimizing this divergence, the algorithm ensures that the imputed distribution matches the target distribution exactly in the limit, without bias. 
Specifically, the algorithm begins by initialization in which  missing entries for any variable are filled with the mean of the observed values, with a small random perturbation added to preserve the marginals. Then, for each iteration, $\hat{\boldsymbol{X}}_L$ and $\hat{\boldsymbol{X}}_K$ are sampled randomly from $\hat{\boldsymbol{X}}$. The Sinkhorn divergence between batches is minimized to obtain  $\mathcal{S}_{\varepsilon}(\mu^*, \nu^*)$. Finally, we update the missing values using standard gradient descent by computing the gradient of the Sinkhorn divergence with respect to the imputed values, i.e.,  $ \nabla_{\Xbm}\mathcal{S}_{\varepsilon}(\mu^*, \nu^*)$. The imputation step is simply moving the missing values in the opposite direction of the gradient, i.e., $\hat{\boldsymbol{X}}_{L \cup  K} \leftarrow \hat{\boldsymbol{X}}_{L \cup  K} - \eta \nabla_{\Xbm}\mathcal{S}_{\varepsilon}(\mu^*, \nu^*)$. Essentially, this step pushes the missing points in batch $\hat{\boldsymbol{X}}_L$ to look more like the distribution of batch $\hat{\boldsymbol{X}}_K$, and vice versa. {The underlying assumption is that the subsets of the dataset, whatever their indices, have similar distributions, which excludes the MNAR case.} This procedure is repeated until convergence. 
}
 
\paragraph*{Takeaway}
\textcolor{black}{ Deep learning has recently emerged as a powerful framework for addressing complex problems involving missing data. Unlike traditional methods, deep models can capture nonlinear dependencies and complex structures, making them particularly effective for high-dimensional and heterogeneous data. MVAE offers a principled generative model, providing coherent joint imputations. However, it is computationally heavy and sensitive to model design. MIWAE improves VAE-based imputations through importance weighting, giving better likelihood estimates and sharper reconstructions, and its extension, not-MIWAE, efficiently handles MNAR data. Besides, GAIN is flexible and can produce realistic imputations with minimal assumptions, but its deep learning architecture requires careful tuning and this is only suited to the MCAR mechanism. OT methods provide geometric interpretability and do not rely on strong parametric assumptions. Nevertheless, they can be computationally costly, and, like GAIN, are not designed handle MNAR data.}

\subsection*{Prediction despite missing values} \label{sec:discriminative}

When the objective is to predict a target value $\ynew$ (e.g., a label) from a vector of new observations $\xnew$, several scenarios must be considered, depending on whether missing values appear in the training set $\Xtrain$, in the new observations $\xnew$ or in both. \textcolor{black}{The recommended method varies depending on the scenario.} The target values in the training set, denoted as $\ytrain$, are observed\footnote{Cases where $\ytrain$ are fully or partially missing can be cast as unsupervised or semi-supervised learning and are not covered in this tutorial. \textcolor{black}{There are relatively few studies that address the unsupervised setting when the dataset $\Xb$ contains missing values. To get started in this context, using the tools presented in our tutorial, we recommend that the reader consult \cite{hunt2003mixture}, which relies on the EM algorithm to handle missing data. The semi-supervised setting is rarely approached from a missing-data perspective. For readers interested in this connection, we recommend \cite{grandvalet2004semi,schmutz2022don}.}}.
\begin{itemize}
\item \textit{Only $\Xtrain$ contains missing values}\label{sec:model_pred_Xtrain}: There may be missing data only in the training set; for example, consider the case of historical data from a sensor network recording various seismic variables. In this case, some readings in the training set ($\Xtrain$) may be missing due to sensor downtime or data corruption. In contrast, the test data ($\xnew$) is clean thanks to modernized and redundant monitoring sensors. In such cases, the objective is to learn $f(\yv|\Xb)$. \textcolor{black}{Several} strategies are available: one can either impute the training data $\Xtrain$, adapt the learning algorithm to handle missing values \textcolor{black}{without using direct imputation of missing values (e.g., by using EM-type algorithms) or by applying naive imputation and accounting for the bias it may introduce};
\item \textit{Both $\Xtrain$ and $\xnew$ contain missing values}\label{sec:model_pred_Xtrain_Xnew}:
When missing values are present in both the training set and the new observations, the key is to be able to make predictions despite incomplete data. The objective becomes estimating $f(\yv|\Xbo)$, and it is not necessary to model the full distribution $f(\Xb)$. In particular, simple imputation methods (e.g., constant imputation) before applying classical algorithm to predict may suffice. \textcolor{black}{An other strategy that we discuss below integrates the management of missing values directly in a decision tree algorithm.}
\end{itemize}
\textcolor{black}{Note that some of these approaches do not explicitly model data distribution $f(\Xb)$ and focus on directly modeling the relationship between inputs and outputs ($f(\yv|\Xb)$ or $f(\yv|\Xbo)$). They can be considered as discriminative methods for missing data.}

\subsubsection*{Predict when only $\Xtrain$ is incomplete}\label{sec:adap_learning}

\textcolor{black}{When there are missing values only in $\Xtrain$, two strategies can be used. The first consists in applying a complex imputation model, ${I}_{\textrm{comp}}(\Xtrain,\ytrain)$ (such as the model-based approaches discussed in the \textit{Imputation} section), to obtain a fully imputed training set. A standard prediction algorithm is then trained on this completed dataset. It should be noted that the imputation model must incorporate information from the target variable $\yv$ in order to capture its relationships with $\Xtrain$ and its potential influence on the missing-data mechanism \cite{vanbuuren2018}.}

\textcolor{black}{
The second strategy consists in estimating a coefficient parameter $ \boldsymbol{\beta} \in \mathbb{R}^p$ despite missing values in some model connecting $\yv$ and $\Xb$. For example, in the linear case, one has $\yv=\Xb^\top\boldsymbol{\beta}+\boldsymbol{\epsilon},$ with $\boldsymbol{\epsilon}\in \mathbb{R}^p$ a noise term. The estimation of $\boldsymbol{\beta}$ performed in the training set can then be used to predict $\ynew=\xnew^\top\boldsymbol{\beta}$. Here, the estimation step can be done by applying adapted learning algorithms for missing values like EM-algorithms (see \textit{Estimation} section) or by naively imputing the missing values and account for this error in a debiased algorithm.
}
More specifically, if the target algorithm $A$ is designed for complete data, the steps are the following
\begin{enumerate}[label=(\roman*)]
    \item A simple imputation method \textcolor{black}{${I}_{\textrm{simp}}(\cdot)$}, such as constant imputation, is applied to obtain a completed \textcolor{black}{training} set ${I}_{\textrm{simp}}(\Xtrain)$.
    \item Algorithm $A$ is then modified to correct for the bias introduced by the imputation. This debiased version of $A$ is then applied to the imputed data ${I}_{\textrm{simp}}(\Xtrain)$.
\end{enumerate}
This strategy has been mostly studied in the linear regression setting when covariates are missing \cite{loh2011high,sportisse2020debiasing}, i.e.\ $\mathbb{E} \left[ \yv|\Xb \right] = h(\Xb)$, with $h(\cdot)$ a linear function. \textcolor{black}{For clarity, an example of this strategy is given in the context of ridge regression to apply the stochastic gradient descent (SGD) algorithm when there are MCAR values in $\Xtrain$ \cite{sportisse2020debiasing}. The goal is to estimate the coefficient parameter $\boldsymbol{\beta}$ of the linear regression model $y_i=\xv_j^\top\boldsymbol{\beta}+\epsilon_i,$ where $\epsilon_i\in \mathbb{R}$ is a noise term, by minimizing the risk $R(\boldsymbol{\beta})=\mathbb{E}[(\xv_j^\top\boldsymbol{\beta}-y_j)^2/2]$. Without missing values, the chosen direction at iteration $k$ of the algorithm is $\beta_k = \beta_{k-1} - \eta \: {g}_{k}(\beta_{k-1})$, with ${g}_{k}(\beta_{k})$ is an unbiased gradient of the risk, such that $\mathbb{E}[{g}_{k}(\beta_{k})]=\nabla R(\beta_{k})$. Without missing values, 
\begin{equation}\label{eq:gk_withoutNA}
{g}_{k}(\beta_{k-1})=\xv_k (\xv_k^\top\beta_{k-1}-y_i) \quad \textrm{(without \texttt{NA})}.
\end{equation} In presence of missing values (\texttt{NA}), after imputing the missing entries by 0 to obtain a complete dataset ($\Mb\odot \Xb$), the debiased version of the direction at iteration $k$ of the SGD algorithm involves 
\begin{equation}\label{eq:gk_withNA}
\tilde{g}_{k}(\beta_k)=\frac{1}{p}(\mv_k\odot\xv_k)\left( \frac{1}{p}(\mv_k\odot\xv_k)^\top\beta_k-y_{k}\right) -\frac{1-p}{p^2}\mathrm{diag}\left((\mv_k\odot\xv_k)(\mv_k\odot\xv_k)^\top\right)\beta_{k}, 
\end{equation}
with $p$ the probability for a value of being observed. For obtaining this debiased version, one has for example $\mathbb{E}_M[\mv_k\odot \xv_k]=p\xv_k$ because $M$ follows a Bernoulli distribution of parameter $p$ in the MCAR case. To debias the part involving $\xv_k$ in the expression of $g_k$ without missing values \eqref{eq:gk_withoutNA}, one uses $\frac{1}{p}(\mv_k\odot \xv_k)$ instead of $\mv_k\odot \xv_k$ in \eqref{eq:gk_withNA}. We therefore take into account the fact that missing values have been imputed with 0 in $\mv_k\odot \xv_k$.  
}


\subsubsection*{Predict when $\Xtrain$ and $\xnew$ are incomplete}
When there are missing values in both the training data and the new observations, there are two different strategies \cite{josse2024consistency}: (i) the first one is to use a learning algorithm that directly handles missing data, 
and (ii) the second strategy is two-step: first impute to obtain a complete dataset, then train a learning algorithm on the imputed data. \textcolor{black}{A key assumption behind these strategies is that the training set and test set have the same data distribution, but also the same distribution for missing data. This implies, for example, that if the missing values are MCAR in the training set, they will also be MCAR in the test set.}

\begin{figure}
 \begin{tikzpicture}[scale=0.3]

\node (R) at (0, 1) {\bf root};
\node (G) at (-6, -2) {$X_{i,j}> \textrm{threshold}$};
\node (D) at (6, -2) {$X_{i,j}> \textrm{threshold}$};
\node (NA) at (-5, -3) {or {\texttt{NA}}};

\draw[thick,-] (R) -- (D) ;
\draw[thick,-] (R) -- (G) ;

\node (R2) at (0, -6) {\bf root};
\node (G2) at (-5, -9) {$X_{i,j}\leq \textrm{threshold}$};
\node (D2) at (5, -9) {$X_{i,j} > \textrm{threshold}$};
\node (NA2) at (5, -10) {or {\texttt{NA}}};

\draw[thick,-] (R2) -- (D2) ;
\draw[thick,-] (R2) -- (G2) ;
\end{tikzpicture}
\quad
\quad
\quad
\quad
\begin{tikzpicture}[scale=0.3
]

\node (labelXtrain) at (0,2) {$\Xtraino$};
\matrix (Xtrain) [matrix of nodes,
     nodes={draw, minimum width=0.5cm, minimum height=0.7cm, text width=0.5cm, anchor=center},
     below=0mm of labelXtrain
]{
    20 &  |[fill=gray!30]| \textcolor{black}{25} & 30  \\
    14 & 10 & |[fill=gray!30]| \textcolor{black}{12} \\
    6 & |[fill=gray!30]| \textcolor{black}{4} & 2 \\
};

\node (labelXright) at (8.5,2) {$M$ (when added)};
\matrix (Xright) [matrix of nodes,
     nodes={draw, minimum width=0.5cm, minimum height=0.7cm, text width=0.5cm, anchor=center},
     dashed,
     right=0mm of Xtrain
]{
    1 & 0 & 1\\
    1 & 1 & 0 \\
    1 & 0 & 1\\
};

\node (labelYtrain) [below=0.2cm of Xtrain] {$\ytrain$};

\matrix (Ytrain) [matrix of nodes,
     nodes={draw,minimum width=0.5cm, minimum height=0.7cm, text width=0.5cm, anchor=center},
     below=0mm of labelYtrain
]{
    1 & 1 & 0 \\
};

\node (labelXnew) [right=1.4cm of labelXright] {$\xnewo$};
\matrix (Xnew) [matrix of nodes,
     nodes={draw, minimum width=0.5cm, minimum height=0.7cm, text width=0.5cm, anchor=center},
     below=0mm of labelXnew
]{
    |[fill=gray!30]| \textcolor{black}{25} \\
    13 \\
    4 \\
};

\node[draw, thick, fit=(labelXtrain)(Xtrain)(Xright)(Ytrain), inner sep=0.1cm, label=below:Train] {};

\node[draw, thick, fit=(labelXnew)(Xnew), inner sep=0.1cm, label=below:Test] {};

\end{tikzpicture}
\caption{\textcolor{black}{\emph{Left}: Illustration of a split in Missing Incorporated in Attributes (MIA) method. The algorithm chooses the splitting variable $[\Xb]_{i,:}$, the threshold and the optimal direction to send missing values, the left or right side, represented by the graphs above. \emph{Right}: Two-step strategy, when both $\Xtrain$ and $\xnew$ are missing. $\Xtraino$ and $\xnewo$ denote the observed part of $\Xtrain$ and $\xnew$, respectively. The gray cases are missing values, and the black entries are imputed values. For examples, the variable ${[\Xtrain]}_{1,:}$ has been imputed by its mean, that is 25. The same imputation model is used for imputing the first variable ${[\xnew]}_{1}$ in the test set.}}
\label{fig:predict_both}
\end{figure}

\subsubsection*{1) One-step strategy}
One-step often refers to the Missing Incorporated in Attributes (MIA) method to handle missing values in
tree-based methods \cite{twala2008good}. 
In a decision tree without missing data, at each node, the algorithm chooses the splitting variable and the threshold that minimizes the error. When a variable contains missing values, MIA considers missingness itself as informative: for each candidate split, it evaluates not only the threshold but also the optimal direction (left or right) to send missing values \textcolor{black}{(see left panel of Figure \ref{fig:predict_both})}. This allows the model to learn during training the best routing decision for missing data, rather than imputing or discarding them. As a result, MIA integrates missingness directly into the tree structure and can capture patterns where the presence or absence of a value carries the signal. This makes MIA particularly well-suited to MNAR scenarios. 


\subsubsection*{2) Two-step strategy} The two-step strategy, also known as the \textit{impute-then-regress} approach, involves two distinct phases, imputation followed by prediction, 
as follows: 
\begin{enumerate}[label=(\roman*)]
    \item The training set $\Xtrain$ is imputed using a simple imputation method ${I}_{\textrm{simp}}(\cdot)$.
    \item \label{2step:regress}  A predictive model $\hat{d}$ is trained on the imputed dataset \textcolor{black}{$({I}_{\textrm{simp}}(\Xtrain), \ytrain)$}.
    \item The new observations $\xnew$ is predicted using the model from step \ref{2step:regress}, i.e., \textcolor{black}{$\ynew = \hat{d}({I}_{\textrm{simp}}(\xnew))$}.
\end{enumerate}

It has first been emphasized that the same imputation model should be used for both the training data and the new observations \cite{josse2024consistency}. For example, if \textcolor{black}{${I}_{\textrm{simp}}(\cdot)$} corresponds to mean imputation, the mean of each variable $i$ is computed from the training set ${[\Xtrain]}_{i,:}$ and used to impute missing values in both ${[\Xtrain]}_{i,:}$ and ${[\xnew]}_{i}$ \textcolor{black}{(see right panel of Figure \ref{fig:predict_both})}. This requirement highlights the need for an out-of-sample imputation method. It has been shown that simple methods, such as mean imputation, combined with a powerful learner (typically random forests) at step \ref{2step:regress} can be sufficient. More precisely, if the learning algorithm $\hat{d}$ is universally consistent when trained on complete data, it remains consistent within the impute-then-regress framework \cite{josse2024consistency}, \textcolor{black}{even if simple imputation methods are used}.

It is also commonly recommended to include the missing-data pattern $\Mb$ (see \eqref{eq:mask}) as additional input variables when training the predictive model $\hat{d}$, i.e., by concatenating the observed data matrix with the corresponding missing-data pattern \textcolor{black}{(see right panel of Figure \ref{fig:predict_both})}. Intuitively, this corresponds to the case where the missing value itself carries predictive information. For example, if sensor measurements are frequently missing during extreme weather events (e.g., sensor dropout in storms) when the goal is to forecast weather anomalies, this information can be leveraged \cite{perez2022benchmarking, josse2024consistency}. 

Overall, the choice of the imputation method becomes less critical when the predictor is sufficiently flexible and the missing-data pattern is included 
\cite{morvan2024imputation}.

\paragraph*{Takeaway} \textcolor{black}{When the goal is prediction, identifying which dataset contains missing values becomes crucial. If only $\Xtrain$ is incomplete, missing data must be handled carefully: either by applying an advanced imputation technique to obtain a complete dataset, or by using methods that estimate the relationship between $\yv$ and $\Xb$ on the incomplete training set before applying it to the test set. A simple approach consists of imputing the missing values naively and then adapting the estimation procedure to account for the imputation error. When both $\Xtrain$ and $\xnew$ contain missing values, imputation becomes less critical, because the objective is to learn a model that predicts $\yv$ from $\Xbo$, the observed components of $\Xb$, given that future observations will also be incomplete. One possible strategy is to impute missing values in both the training and test sets using the same imputation model, and then apply a powerful learning algorithm on the completed data. Finally, note that this tutorial does not address scenarios involving distributional shifts between the training and test sets.}

\color{black}

\section*{Conclusion and perspectives}
This tutorial has provided a comprehensive treatment and unifying perspective of modern approaches organized around three core tasks---\textit{imputation},
\textit{estimation}, and \textit{learning}---bridging classical statistical frameworks with contemporary algorithmic innovations.  

\noindent \textbf{Key insights.}
Throughout this tutorial, we have emphasized modern challenges where new methods and algorithms handling missing data have flourished. Effective handling of missing data in signal processing and machine learning requires a paradigm shift from simple, generic imputation to context-aware, model-based strategies that explicitly integrate domain-specific characteristics of the underlying signal: robustness to irregular and heterogeneous data through elliptically symmetric distributions and regularized optimization that maintain performance despite outliers and heavy tails; exploitation of low-dimensional structure via nuclear norm minimization, probabilistic PCA, and graph-regularized recovery that enables both accurate imputation and computational efficiency in high dimensions; and the use of recent learning approaches, such as deep-learning models, for instance variational autoencoders, which offer greater flexibility for addressing complex missing data problems (including MNAR data), along with specific prediction strategies to be adapted depending on the case of missingness.

\noindent
\textbf{Takeaway for practitioners.} When tackling missing data in signal processing or machine learning, the adopted approach should align with both the nature of the data (data matrix, graph signal, time series) and the analytical objective, whether it’s imputation, parameter estimation, or predictive modeling (see each section's \textit{Takeaway}). A good start is to characterize the quantity of missing data and the missingness mechanism, as it fundamentally influences method selection. Specifically, determine whether the missingness is MCAR, MAR, or MNAR. Each mechanism imposes distinct assumptions and constraints on the validity of downstream analyses.

Once the analytical objective is chosen, key trade-offs need to be taken into account for choosing the adapted methods. For imputation, model-based and deep learning methods offer robustness and flexibility for non-ignorable mechanisms but often require more resources. For parameter estimation, likelihood-based methods are well-suited for structured problems as covariance estimation, subspace tracking, or graph learning, offering rigorous uncertainty quantification and robust handling of missing data. However, they often face scalability issues in high dimensions. For prediction tasks, missing values in the training set require careful handling, either through non-naive imputation or by explicitly modeling the link between predictors and response under missingness. When both training and test sets contain missing values, a simple imputation combined with a strong prediction algorithm usually suffices. 

\noindent \textbf{Future directions.}
Numerous promising research directions, tied to the methods discussed in this tutorial, remain unexplored. \textit{Structure-aware approaches} should explore doubly robust algorithms maintaining validity under misspecification and incorporating MNAR relationships grounded in domain-specific physical models. \textit{Graph-based methods} require unified frameworks for MNAR graph learning, scaling to massive streaming networks, and handling heterogeneous graphs with mixed node types. \textit{Uncertainty quantification} demands principled frameworks---leveraging self-supervised and semi-supervised settings, modeling input uncertainty from conflicting sensors, and adapting conformal prediction to missing data---essential for high-stakes applications. Finally, \textit{computational scalability} of deep generative models and stochastic EM variants for incomplete high-dimensional streaming data and large-scale multimodal data remains an important practical challenge.


\color{black}

\bibliographystyle{IEEEtran}

\bibliography{main}
%
%
%
%
%
\section*{Biographies}
\vspace{-2cm}
\begin{IEEEbiographynophoto}{Alexandre Hippert-Ferrer} obtained the Ph.D. degree in remote sensing from LISTIC Laboratory, University Savoie Mont-Blanc, Annecy, France, in 2020. Since 2022, he has been an Associate Professor in machine learning and remote sensing with IGN/ENSG, Université Gustave Eiffel, Champs-sur-Marne, France. His research interests include machine learning and statistical estimation with remote sensing data, including incomplete data.
\end{IEEEbiographynophoto}

\vspace{-2.2cm} 

\begin{IEEEbiographynophoto}{Aude Sportisse}   obtained a Ph.D. degree in statistics from LPSM Laboratory, Sorbonne University, in 2021. Since 2024, she is a CNRS researcher at LIG, Grenoble, France. Her research interests include missing values, semi-supervised learning, low-rank models and clustering.
\end{IEEEbiographynophoto}

\vspace{-2.2cm} 

\begin{IEEEbiographynophoto}{Amirhossein Javaheri} is currently pursuing the dual Ph.D. program in electronic and computer engineering with Sharif University of Technology and The Hong Kong University of Science and Technology. His current research interests include graph learning and signal processing, optimization, machine learning, and data analytics.
\end{IEEEbiographynophoto}

\vspace{-2.2cm} 

\begin{IEEEbiographynophoto}{Mohammed Nabil El Korso} obtained his Ph.D. degree from Paris-Sud XI University, in 2011. Currently, he is Professor at Paris Saclay University. His research interests include robust statistical signal processing, statistical analysis with missing values, estimation with mixed effects models with application to radio-interferometry, SAR and array processing. 
\end{IEEEbiographynophoto}

\vspace{-2.2cm} 

\begin{IEEEbiographynophoto}{Daniel P. Palomar} received the Ph.D. degrees in electrical engineering from the Technical University of Catalonia (UPC), Barcelona, Spain in 2003.  He is a Professor with the Department of Electronic and Computer Engineering and the Department of Industrial Engineering and Decision Analytics at The Hong Kong University of Science and Technology (HKUST), Hong Kong, where he joined in 2006. His current research interests include applications of optimization theory, graph methods, and signal processing in financial systems and big data analytics.
\end{IEEEbiographynophoto}

\end{document}